\newif\ifreport\reporttrue
\newif\iffinal\finalfalse

\ifreport
\documentclass[10pt,journal]{IEEEtran}
\else
\documentclass[10pt,draftcls,onecolumn]{IEEEtran}
\fi

\usepackage{amsfonts}
\usepackage{amsthm}
\usepackage{mathrsfs}
\usepackage{graphicx,epsfig,amssymb,amsmath,url,latexsym}
\usepackage[lined, boxed, linesnumbered, commentsnumbered, ruled]{algorithm2e} %vlined
\usepackage{subfigure}
\usepackage[square,sort&compress,numbers]{natbib}
\usepackage{bm}
\usepackage{multirow}
\usepackage{color}
\usepackage{enumerate}
\usepackage{comment}
\usepackage{setspace}

\newtheorem{theorem}{Theorem}
\newtheorem{lemma}{Lemma}

\newtheorem{definition}{Definition}

\newtheorem{statement}{Statement}
\theoremstyle{remark}
\newtheorem{remark}{Remark}

\begin{document}
% paper title

%\title{Scaling Law of a Moving Window Network Coding Strategy in Reliable Multicast}
\title{Constant Delay and Constant Feedback Moving Window Network Coding for Wireless Multicast: Design and Asymptotic Analysis}

\author{
Fei Wu, Yin Sun, Yang Yang, Kannan Srinivasan, and Ness B. Shroff

\thanks{Fei Wu, Yin Sun, and Yang Yang are with the Department of ECE, The Ohio State University, Columbus, OH, 43210 (e-mail: wuff.gewuer@gmail.com, sunyin02@gmail.com, yang.1267@osu.edu).}
\thanks{Kannan Srinivasan is with the Department of CSE, The Ohio State University, Columbus, OH, 43210 (e-mail: kannan@cse.ohio-state.edu).}
\thanks{Ness B. Shroff is with the Departments of ECE and CSE, The Ohio State University, Columbus, OH, 43210 (e-mail: shroff.11@osu.edu).}
}

\maketitle

%\baselineskip=24pt

%\IEEEcompsoctitleabstractindextext{
\begin{abstract}
A major challenge of wireless multicast is to be able to support a
large number of users while simultaneously maintaining low delay and
low feedback overhead. In this paper, we develop a joint coding and
feedback scheme named Moving Window Network Coding with Anonymous
Feedback (MWNC-AF) that successfully addresses this challenge. In
particular, we show that our scheme simultaneously achieves both a
constant decoding delay and a constant feedback overhead,
irrespective of the number of receivers $n$, without sacrificing
either throughput or reliability. We explicitly characterize the
asymptotic decay rate of the tail of the delay distribution, and
prove that transmitting a fixed amount of information bits into the
MWNC-AF encoder buffer in each time-slot (called ``constant data
injection process'') achieves the fastest decay rate, thus showing
how to obtain delay optimality in a large deviation sense. We then
investigate the average decoding delay of MWNC-AF, and show that
when the traffic load approaches the capacity, the average decoding
delay under the constant injection process is at most one half of
that under a Bernoulli injection process. In addition, we prove that
the per-packet encoding and decoding complexity of MWNC-AF both
scale as $O(\log n)$, with the number of receivers $n$. Our
simulations further underscore the performance of our scheme through
comparisons with other schemes and show that the delay, encoding and
decoding complexity are low even for a large number of receivers,
demonstrating the efficiency, scalability, and ease of
implementability of MWNC-AF.

\end{abstract}

%}
\maketitle

\section{Introduction}
\label{sec:introduction}

Wireless multicast has numerous applications: wireless IPTV,
distance education, web conference, group-oriented mobile commerce,
firmware reprogramming of wireless devices, etc,
\cite{Luby2008,learning,reprogram}. However, in reality, there are
only a few deployments. A major challenge that wireless multicast
techniques have so far not been able to overcome is to achieve low
delay without incurring a large amount of feedback. In the
literature, there are two categories of multicast coding strategies.
The first category focuses on batch-based coding schemes, e.g.,
random linear network coding (RLNC) \cite{ho2006random}, LT codes
\cite{luby2002}, and Raptor codes \cite{shokrollahi2006raptor}. In
these schemes, the transmitter sends out a linear combination
generated from a batch of $B$ data packets in each time-slot. A new
batch of packets cannot be processed until all the receivers have
successfully decoded the previous packet batch. This approach has a
low feedback overhead: one bit of acknowledgment (ACK) is sufficient
to signal the decoding fate of an entire batch. However, with a
fixed batch size, the achievable throughput decreases with the
number of receivers $n$. To maintain a fixed throughput, the batch
size $B$ needs to grow on the order of $O(\log n)$
\cite{swapna,yang2012throughput}. As the batch size $B$ increases,
the decoding delay also grows as $O(\log n)$. Thus, such schemes
achieve low feedback overhead at the cost of high decoding delay.

The second category of studies are centered on an incremental
network coding design\footnote{They are also referred as online or
adaptive network coding in the literatures.}, e.g.,
\cite{towards_ACK,kumar2008arq,sundararajan2009feedback,barros2009effective,parastoo2010optimal,li2011capacity,sorour2010minimum,playback_delay,three_schemes,Ton13,pimrc12,temple13,sundararajan2009network,lin2010slideor,gesture,block_feedback},
where the data packets participate in the coding procedure
progressively. Therefore, the receivers that have decoded old
packets can have early access to the processing of new data packets,
instead of waiting for all the other receivers to decode the old
packets. The benefit of this approach is low decoding delay. Some
studies have even shown a constant upper bound of decoding delay for
any number of receivers, when the encoder is associated with a
Bernoulli packet injection process
\cite{sundararajan2009feedback,Ton13}. However, these schemes need
to collect feedback information from all receivers, and the total
feedback overhead increases with the number of receivers $n$. Thus,
these incremental-coding schemes achieve low delay, but at the cost
of high feedback overhead.

{\bf Can we achieve the best of both worlds?} This paper develops a
joint coding and feedback scheme called Moving Window Network Coding
with Anonymous Feedback (MWNC-AF) that achieves the delay
performance of incremental-coding techniques without requiring the
feedback overhead to scale with the number of receivers, as in the
batch-based coding techniques. Hence, it indeed shows that the best
of both worlds is achievable. We present a comprehensive analysis of
the decoding delay, feedback overhead, encoding and decoding
complexity of MWNC-AF. The contributions of this paper are
summarized as follows:

\begin{itemize}

\item We develop a joint coding and feedback scheme called MWNC-AF, and show that
MWNC-AF achieves both a constant decoding delay and a constant
feedback overhead\footnote{By constant delay and constant feedback
overhead, we mean that the delay experienced by any receiver and the
overall feedback overhead of all receivers are both independent of
the number of receivers $n$.}, irrespective of the number of
receivers $n$, without sacrificing either throughput or reliability.

\item
We investigate how to control the data injection process at the
encoder buffer to reduce the decoding delay of MWNC-AF. To that end,
we explicitly characterize the asymptotic decay rate of the tail of
the decoding delay distribution for any \emph{i.i.d.} data injection
process. We show that injecting a constant amount of information
bits into the encoder buffer in each time-slot (called ``constant
data injection process'') achieves the fastest decay rate, thus
showing how to obtain delay optimality in a large deviation sense.
(Theorem \ref{theorem1})

\item We derive an upper bound of the average decoding delay for MWNC-AF under the constant data injection process. As the traffic load approaches capacity,
this upper bound is at most one half of the average decoding delay
achieved by a Bernoulli data injection process. (Theorem
\ref{theorem2})

\item For the constant data injection process, we prove that the average encoding complexity of MWNC-AF is of the form $\frac{1}{\eta}\log n + o(\log n)$ for sufficiently large $n$, and the value of the pre-factor $\eta$ is attained as a
function of the channel statistics and the injection rate. For any
$n$, we also characterize the asymptotic decay rate of the tail of
the encoding complexity distribution. (Theorem \ref{theorem3})

\item For the constant data injection process, we prove that the average decoding complexity of MWNC-AF per data packet is also of the form $\frac{1}{\eta}\log n + o(\log n)$ for sufficiently large $n$, and the pre-factor $\frac{1}{\eta}$ is the same as that of the average encoding
complexity. (Theorem \ref{theorem4})

%We prove that both the average encoding and decoding complexity of this MWNC transmission scheme grows as $O(\log n)$ when the number of receivers $n$ increases.

%Our simulation results show that, for the case of $n = 1024$ receivers, the average encoding and decoding complexity of this MWNC transmission scheme is less than 50 operations. Therefore, this MWNC transmission scheme has low encoding and decoding complexity for practical values of $n$.

\end{itemize}

The rest of this paper is organized as follows. In Section
\ref{related_work}, we introduce some related work. In Section
\ref{sec:system_model}, we describe the system model and present our
MWNC-AF transmission design. In Section \ref{sec:scaling law}, we
analyze the decoding delay, encoding complexity, and decoding
complexity of the MWNC-AF transmission design. In Section
\ref{sec:simulations}, we use simulations to verify our theoretical
results. Finally, in Section \ref{sec:conclusion}, we conclude the
paper.

\section{Related Work}\label{related_work}
Batch-based rateless codes can generate a potentially unlimited
stream of coded packets from a fixed batch of data packets. The
coded packets can be generated on the fly, as few or as many as
needed \cite{luby2002}. Examples of Batch-based rateless codes
includes random linear network coding (RLNC) \cite{ho2006random}, LT
codes \cite{luby2002}, and Raptor codes
\cite{shokrollahi2006raptor}. RLNC\footnote{By RLNC, we refer to the
specifications in \cite{ho2006random,coding_gain,swapna}.} is the
simplest rateless codes, which can achieve near-zero communication
overhead. However, the decoding complexity of RLNC is high
\cite{MacKay05} for large block size $B$. LT codes and Raptor codes
were proposed to reduce the decoding complexity. In particular,
Raptor codes can achieve constant per-packet encoding and decoding
complexity. One benefit of batch-based rateless codes is low
feedback overhead \cite{xiao2012reliable}. A feedback scheme was
proposed in \cite{SMART} for RLNC, which has a constant overhead
independent of the number of receivers. However, these schemes have
poor delay performance when the number of receivers $n$ is large.
Recent analyses have shown that, to maintain a fixed throughput, the
batch size in these schemes needs to grow with respect to the number
of receivers $n$, which results in a long decoding delay
\cite{swapna,yang2012throughput}. Scheduling techniques have been
developed to optimize the tradeoff between the batch size and
throughput under limited feedback for finite $n$
\cite{Sun2013,Sun2014}. However, it is difficult to maintain a low
decoding delay for large $n$, unless resorting to novel coding
designs.

In recent years, a class of incremental network coding schemes,
e.g.,
\cite{towards_ACK,kumar2008arq,sundararajan2009feedback,barros2009effective,parastoo2010optimal,li2011capacity,sorour2010minimum,playback_delay,three_schemes,Ton13,pimrc12,temple13,sundararajan2009network,lin2010slideor,gesture,block_feedback}
are developed to resolve the long decoding delay of rateless codes.
In these designs, the data packets participate in the coding
procedure progressively. Among this class, an instantly decodable
network coding scheme was proposed in
\cite{parastoo2010optimal,li2011capacity}, where the number of
receivers that can be effectively supported is maximized under a
zero decoding delay constraint. In order to accommodate more
receivers, the zero decoding delay constraint was relaxed in
\cite{sorour2010minimum}. Nonetheless, these schemes cannot support
a large number of receivers.

A number of ARQ-based network coding schemes are proposed since the
seminal work \cite{kumar2008arq,sundararajan2009feedback}, which can
potentially reduce the decoding delay and support a large number of
receivers. In \cite{kumar2008arq,sundararajan2009feedback}, the
desired packet of each receiver is acknowledged to the transmitter,
such that the transmitted packet is a linear combination of the
desired packets of all receivers. Without appropriate injection
control, this scheme results in unfair decoding delay among the
receivers with different packet erasure probabilities. A
threshold-based network coding scheme was proposed in
\cite{barros2009effective} to resolve this fairness issue, at the
cost of some throughput loss. A dynamic ARQ-based network coding
scheme was proposed in \cite{Ton13}, which can achieve noticeable
improvement in the throughput-delay tradeoff performance.
Interestingly, when associated with a Bernoulli packet injection
process, the average decoding delay of ARQ-based network coding is
upper bounded by a constant\footnote{When the number of receivers
$n$ is small, the average decoding delay in
\cite{sundararajan2009feedback,Ton13} can be substantially smaller
than the upper bound. However, when there are a large number of
receivers, the average decoding delay in
\cite{sundararajan2009feedback,Ton13} is very close to the upper
bound, as shown in \cite{Ton13}.} independent of the number of
receivers $n$ \cite{sundararajan2009feedback,Ton13}. However, these
schemes require explicit feedback from each receiver, and thus their
feedback overhead scales up with the number of receivers $n$. A
generalization of ARQ-based network coding was the moving window
network coding (MWNC), which was first proposed in
\cite{sundararajan2009network} to make network coding compatible
with the existing TCP protocol. The MWNC scheme was also employed in
multihop wireless networks to improve the throughput of
opportunistic routing \cite{lin2010slideor} and support multiple
multicast sessions \cite{gesture}. However, in these designs, the
movement of the encoding window requires the ACK from all receivers,
and thus the feedback overhead scales up with the network size.

Recently, the first author proposed an anonymous feedback scheme for
MWNC \cite{pimrc12}, which can achieve a constant feedback overhead
for any number of receivers $n$. However, this feedback scheme
assumed that all receivers are within a short range of each other
and can communicate with one another, which may introduce the
well-known hidden terminal problem in practical systems. In
addition, the window size of the MWNC scheme was fixed in
\cite{pimrc12}, which leads to a throughput degradation as the
number of receivers $n$ grows up. Another low-overhead feedback
scheme was proposed in \cite{temple13} for ARQ-based network coding,
where only the leading and tail receivers feed back messages to the
transmitter. However, it was not discussed in \cite{temple13}
whether their scheme can achieve constant decoding delay for any
number of receivers. To the extent of our knowledge, no previous
scheme exists that can simultaneously guarantee constant decoding
delay and constant feedback overhead as the number of receivers $n$
grows, without sacrificing the throughput and reliability of
wireless multicast.

\section{System Model}\label{sec:system_model}

\subsection{Channel Model}
We consider a broadcast packet erasure channel with one transmitter
and $n$ receivers, where the transmitter needs to send a stream of
common information to all the receivers.

We assume a time-slotted system. In each time-slot, the transmitter
generates one coded packet and broadcasts it to all the receivers.
The channel from the transmitter to the receiver $i$ in time-slot
$t$ is denoted as $c_i[t]$, where \ifreport
\begin{align}
\!\!\!\!c_i[t]=\left\{
\begin{array}{ll}
1&\text{if a coded packet is successfully}\\
&\text{received by receiver $i$ at time-slot $t$;}\\
0&\text{otherwise.}
\end{array}\right.\label{cit_def}
\vspace{0.3cm}
\end{align}
\else
\begin{align}
\!\!\!\!c_i[t]=\left\{
\begin{array}{ll}
1&\text{if a coded packet is successfully received by receiver $i$ at time-slot $t$;}\\
0&\text{otherwise.}
\end{array}\right.\label{cit_def}
\vspace{0.3cm}
\end{align}
\fi We assume that $c_i[t]$ is {\it i.i.d.} across time-slots, and
define $\gamma_i\triangleq \mathbb{P}\left(c_i[t]=1\right)$. Then,
the capacity of this broadcast channel is $\inf_{1\le i\le n}
\gamma_i$ packets per time-slot.

It is assumed that on the feedback channel, the transmitter and each
receiver can overhear each other, but the receivers may not overhear
each other. Since all receivers are within the one-hop transmission
range of the transmitter and in practice the feedback signals are
usually sent at a much lower data rate than the normal data packet,
similar to
\cite{towards_ACK,kumar2008arq,sundararajan2009feedback,barros2009effective,parastoo2010optimal,li2011capacity,sorour2010minimum,playback_delay,three_schemes,Ton13,pimrc12,temple13,sundararajan2009network,lin2010slideor,gesture,SMART},
we assume that the feedback signals can be reliably detected.

%Our analysis also applies to the
%systems with asymmetric packet erasure probabilities for different
%users; see Section \ref{} for the details.
%Under this setting, it is easy to see that

%where $c_i[t] = 1$ represents that the coded packet is successfully received by receiver $i$, and $c_i[t] = 0$ represents that the coded packet is erased at receiver $i$. The packet reception probability of receiver $i$ is $\Pr\{c_i[t] = 1\} = \gamma_i$. We assume that $c_i[t]$ is independent across the receivers and \emph{i.i.d.} across time slots.

%In the next subsection, we propose a new broadcast scheme, in which the achievable throughput is a control parameter that can be made arbitrarily close to $\gamma$.

\subsection{Multicast Transmission Design}
%{\red very careful to state here.}Inspired by
%\cite{sundararajan2009network}, we now introduce a  with an anonymous feedback scheme,
%which is able to achieve constant decoding delay and constant
%feedback overhead for any number of receivers $n$.

We propose a multicast transmission scheme called moving window
network coding with anonymous feedback (MWNC-AF). This scheme
achieves a constant decoding delay and a constant feedback overhead
for any number of receivers $n$.
%We will show that the average decoding delay and feedback overhead of this scheme remains constant for any number $n$ of the receivers, and thereby achieves the optimal scaling law.

%where an encoding block of packets is maintained, but different from
%traditional approaches, the block evolves consecutively over time
%according to prescribed parameters and the feedback from receivers.

\subsubsection{Encoder}
Assume that the transmitter is infinitely backlogged, and that
$\tilde{a}[t]$ bits are injected to the encoder buffer from the
backlog at the beginning of time-slot $t$. The bits received by the
encoder are assembled into packets of $L$ bits. Let us define
$a[t]={\tilde{a}[t]}/{L}$, which is a rational number. We assume
that $a[t]$ is \emph{i.i.d.} across time-slots with mean $\lambda
\triangleq \mathbb{E}\left[a[t]\right]$. Then, the number of packets
that the encoder has received up to the beginning of time-slot $t$
is $A[t]$, i.e.,
\begin{align}
A[t]=\sum_{\tau=1}^t a[\tau].\label{At_def}
\end{align}
We note that only fully assembled packets can participate the
encoding operation. The number of fully assembled packets up to the
beginning of time-slot $t$ is $\left\lfloor A[t]\right\rfloor$,
where $\lfloor y \rfloor$ is the maximum integer no greater than
$y$.

Let $Z[t]\in\mathbb{N}$ denote the number of packets that have been
removed from the encoder buffer by the end of slot $t$. The
evolution of $Z[t]$ will be explained in Section~\ref{sec:feedback},
along with the anonymous feedback scheme. The coded packet $x[t]$ in
time-slot $t$ is generated by
\begin{align}
x[t]=\sum_{m=Z[t-1]+1}^{\left\lfloor
A[t]\right\rfloor}\alpha_{t,m}\times p_{m},\label{rand_linear_comb}
\end{align}
where $p_m$ denotes the $m^{\text{th}}$ assembled packet of the
encoder, ``$\times$'' is the product operator on a Galois field
$GF(2^q)$, and $\alpha_{t,m}$ is randomly drawn according to a
uniform distribution on $\{GF(2^q)\}\backslash\{0\}$.\footnote{The
bit-size $L$ of each packet is a multiple of $q$.} The values of
$\left\lfloor A[t]\right\rfloor$ and $Z[t-1]$ are embedded in the
packet header of $x[t]$. In addition, $\{\alpha_{t,m}\}$ are known
at each receiver by feeding the same seed to the random number
generators of the transmitter and all the receivers.
%{\color{red}We make the usual assumption that the parameters $\left\lfloor A[t]\right\rfloor$, $Z[t]$, and $\alpha_{t,m}$ are known to the receivers that successful retrieve $x[t]$, by adding packet header and synchronizing the random number generators.}

Let $W[t]$ denote the number of packets that participate in the
encoding operation of $x[t]$ in time-slot $t$, which is called
\emph{encoder queue length} or \emph{encoding window size} in this
paper. According to Equation~\eqref{rand_linear_comb}, $W[t]$ is
determined by
\begin{align}\label{eq_W}
W[t] &= \left\lfloor A[t]\right\rfloor-Z[t-1].
\end{align}

\subsubsection{Decoder}
\begin{figure}[t]
\centering
  \includegraphics[width=3.5in]{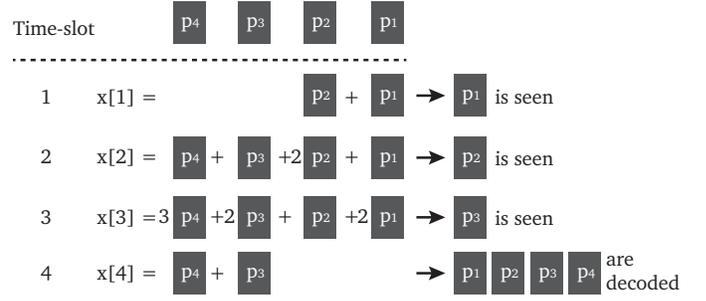}
  %\vspace{-0.25cm}
  \caption{An example for the decoding procedure of MWNC-AF.}
  \vspace{-0.5cm}
  \label{fig:seen}
\end{figure}
To facilitate a clear understanding of the decoding procedure, we
restate the definition of {\it a user seeing a packet} that was
originally described in \cite{kumar2008arq}.

\begin{definition} (Seeing a packet)
We say that a receiver has ``seen'' a packet $p_m$, if it has enough
information to express $p_m$ as a linear combination of some packets
$p_{m+1}, p_{m+2}, \cdots$ with greater indices.
\end{definition}

We first use the example illustrated in Fig.~\ref{fig:seen} to
explain the decoding procedure. In this example, the coded packets
$x[1]$, $x[2]$, $x[3]$, and $x[4]$ are successfully delivered to a
certain receiver in time slots 1-4, respectively. In time-slot 1,
packet $p_1$ is ``seen'' at the receiver, because it can be
expressed as
$$p_1=x[1]-p_2.$$
Similarly, in time slots 2-4, packets $p_2$, $p_3$, and $p_4$ are
``seen'' one by one, because they can be expressed as
\begin{eqnarray}
&&p_2=x[2]-x[1]-p_4-p_3,\nonumber\\
&&p_3=x[3]-x[2]+x[1]-2p_4,\nonumber\\
&&p_4=-x[4]+ x[3]-x[2]+x[1].\nonumber
\end{eqnarray}
Now, packet $p_4$ is immediately decoded, because $x[1]$, $x[2]$, $x[3]$, and $x[4]$ are available at the receiver. Once $p_4$ is decoded, it can be substituted backwards to decode $p_3$, $p_2$, and $p_1$ one by one.

%Substituting $p_4$ into the expression of $p_3$, we can decode $p_3$. Using such a backward substitution procedure, we can also decode $p_2$ and $p_1$.

%Let $p_1,\cdots,p_{m-1}$ be the packets that receiver $i$ has seen up to slot $t$.

Let $S_i[t]\in\mathbb{N}$ be the number of packets that receiver $i$
has ``seen'' by the end of time-slot $t$. Define a \emph{virtual
decoder queue}
\begin{align}
Q_i[t]=A[t]-S_i[t]\label{decoder_queue}
\end{align}
for each receiver $i$. Then, $\left\lfloor Q_i[t] \right\rfloor$ is
the number of ``unseen'' packets at receiver $i$ at the end of
time-slot $t$.

The decoding procedure of receiver $i$ is described as follows:

At the beginning of time-slot $t$, receiver $i$ has seen the packets
$p_1,\cdots,p_{S_i[t-1]}$. Suppose $c_i[t] =1$, which implies that
packet $x[t]$ is successfully delivered to receiver $i$ in time-slot
$t$. If $A[t]-S_i[t-1]\ge 1$, the packets participated in generating
$x[t]$ contains at least one ``unseen'' packet $p_{S_i[t-1]+1}$.
Receiver $i$ eliminates the ``seen'' packets
$p_1,\cdots,p_{S_i[t-1]}$ from the expression of $x[t]$ in
Equation~\eqref{rand_linear_comb} of $x[t]$, to obtain an expression
of $p_{S_i[t-1]+1}$. If the field size $2^q$ is sufficiently large,
then with high probability, packet $p_{S_i[t-1]+1}$ can be expressed
as a linear combination of the packets $p_{S_i[t-1]+2},
p_{S_i[t-1]+3}, \cdots$ with greater indices. In other words, packet
$p_{S_i[t-1]+1}$ is ``seen'' in time-slot $t$. Therefore, the value
of $S_i[t]$ can be updated by
%\begin{eqnarray}
%S_i[t]=\left\{
%\begin{array}{ll}
%S_i[t-1]+c_i[t], & \mbox{\normalfont if}\: \left\lfloor A[t]\right\rfloor-S_i[t-1]\ge 1; \\
%S_i[t-1], &\mbox{\normalfont otherwise.}
%\end{array}
%\right.\!\!\!\label{S_def}
%\end{eqnarray}
\begin{eqnarray}
S_i[t]=S_i[t-1]+c_i[t] 1_{\left\{A[t]-S_i[t-1]\ge 1\right\}}, \label{S_def}
\end{eqnarray}
where ${1}_A$ is the indicator function of event $A$.

If
\begin{eqnarray}\label{def_decoding_moment}
%\left\lfloor A[t]\right\rfloor=S_i[t],
\left\lfloor A[t]\right\rfloor=S_i[t] \text{~or~equivalently~}
\left\lfloor Q_i[t] \right\rfloor = 0,
\end{eqnarray}
i.e., receiver $i$ has ``seen'' all the packets that participated in
the encoding operation of $x[t]$, then receiver $i$ can decode
packet $p_{S_i[t]}$. Once $p_{S_i[t]}$ is decoded, it can be
substituted backwards to sequentially decode $p_{S_i[t]-1}$,
$p_{S_i[t]-2}$, $\cdots$, for all ``seen'' packets.
\subsubsection{Anonymous Feedback}\label{sec:feedback}
According to the decoding procedure, if a packet $p_m$ is ``unseen''
at some receiver $i$, it cannot be removed from the encoder buffer.
Because, otherwise, receiver $i$ will never be able to ``see''
packet $p_m$ or decode it. In order to ensure reliable multicast,
the departure process $Z[t]$ of the encoder buffer should satisfy
\begin{eqnarray}
Z[t] \leq \min_{1\leq i\leq n} S_i[t].\label{Z_S0}
\end{eqnarray}

%In order to ensure reliable multicast and , the departure process $Z[t]$ of the encoder buffer should satisfy

%such that all the unseen packets are kept in the encoder buffer.
%
%We now provide a beacon-based anonymous feedback scheme, which can guarantee
%\begin{align}
%Z[t] = \min_{1\leq i\leq n} S_i[t].\label{Z_S}
%\end{align}
%By this,

\begin{algorithm}[h]
\setstretch{1.0}
%\dontprintsemicolon
%\color{blue}
\SetKwIF{If}{ElseIf}{Else}{if}{then}{else if}{else}{endif}
\SetKwInput{AtNodeI}{Feedback procedure of receiver $i$}
\SetKwInput{AtBaseStation}{Reaction procedure of the transmitter}
\Indentp{-1em}\AtNodeI{}\Indentp{1em} ${Z}_i[0]:=0$\; $S_i[0]:=0$\;
\For{time slot $t=1:\infty$}{
    - - - - - - - Data sub-slot - - - - - - - - - -\\
    Receive coded packet $x[t]$\;
%   \uIf{A coded packet is received at receiver $i$}{
%       set $S_i[t]=S_i[t-1]+1$\;
%   }\Else{
%       set $S_i[t]=S_i[t-1]$\;
%   }
    Update $S_i[t]$ according to Equation~\eqref{S_def}\;
%set $I^\text{lead}_i=I^\text{lead}_i+1$\;
    - - - - - - - Beacon sub-slot - - - - - - - - \\
    \uIf{$S_i[t]=Z_i[t-1]$}{
        Send out a beacon signal\;
        ${Z}_i[t]:={Z}_i[t-1]$\;
    }\Else{
    Detect beacon signal\;
    %- - - - - Beacon sub-slot 2 - - - - - - - -\\
    %\Else{
    %    Detect beacon signal\;
        \uIf{no beacon signal is detected}{
                ${Z}_i[t]:={Z}_i[t-1]+1$\;
        }\Else{
                ${Z}_i[t]:={Z}_i[t-1]$\;

        }
    }

%   - - - - - After the beacon slot of $t$ - - - - - \\
%   \uIf{no beacon signal is transmitted or detected in the beacon slot}{
%       set ${Z}_i[t]={Z}_i[t-1]+1$\;
%   }\Else{
%       set ${Z}_i[t]={Z}_i[t-1]$\;
%   }
} \Indentp{-1em}\AtBaseStation{}\Indentp{1em} $Z[0]:=0$\; \For{time
slot $t=1:\infty$}{
    - - - - - - - Data sub-slot - - - - - - - - - -\\
    Send coded packet $x[t]$\;
    - - - - - - - Beacon sub-slot - - - - - - - - -\\
    Detect beacon signal\;
    %- - - - - - - Beacon sub-slot 2 - - - - - - - -\\
    \uIf{beacon signal is detected}{
        Send out a beacon signal\;
        $Z[t]:=Z[t-1]$\;
    }\Else{
        $Z[t]:=Z[t-1]+1$\;
        Remove the oldest packet from the encoder\;
    }
} \caption{Beacon-based Anonymous Feedback} \label{feedbackalg}
\end{algorithm}%\vspace{-0.3cm}

We now provide a beacon-based anonymous feedback scheme, provided in
Algorithm~\ref{feedbackalg}. In this algorithm, receiver $i$
maintains a local parameter $Z_i[t]$, which is synchronized with
$Z[t]$ at the transmitter through beacon signaling. Each time-slot
is divided into a long data sub-slot and a short beacon sub-slot. In
the data sub-slot, the transmitter broadcasts a data packet to all
the receivers. Then, $S_i[t]$ is updated according to
Equation~\eqref{S_def}. In the beacon sub-slot, if receiver $i$
finds that $S_i[t]=Z_i[t-1]$, it sends out a beacon signal in the
common feedback channel, requesting the transmitter not to remove
the oldest packet in the encoder buffer. If the transmitter has
detected the beacon signal (from one or more receivers), the
transmitter will broadcast a beacon signal instantly within the same
beacon sub-slot, and no packet will be removed from the encoder
buffer, i.e.,
\begin{align}\label{Zt_2}
Z[t] =Z[t-1].
\end{align}
In the beacon sub-slot, the transmitter serves as a relay for the
beacon signal. This second beacon transmission guarantees that
receivers that are hidden from each other can still detect each
other¡¯s beacon signal. If the transmitter has detected no beacon
signal, it will remove the oldest packet in the encoder buffer,
i.e.,
\begin{align}\label{Zt_1}
Z[t] =Z[t-1]+1.
\end{align}
By detecting the existence of beacon signal in the beacon-sub-slot,
each receiver synchronizes $Z_i[t]$ with $Z[t]$. A key benefit of
this anonymous feedback scheme is that \textbf{its overhead (i.e.,
one short beacon sub-slot) is constant for any number of receivers
$\textbf{n}$}.

%
% If the transmitter has detected no beacon signal, it will remove the oldest packet in the encoder buffer and transmit no signal in beacon sub-slot 2. Otherwise, if the transmitter has detected the beacon signal (from one or more receivers), no packet will be removed from the encoder buffer and the transmitter will broadcast a beacon signal to all the receivers in beacon sub-slot 2.

\begin{figure}[t]
\centering
  \includegraphics[width=3.2in]{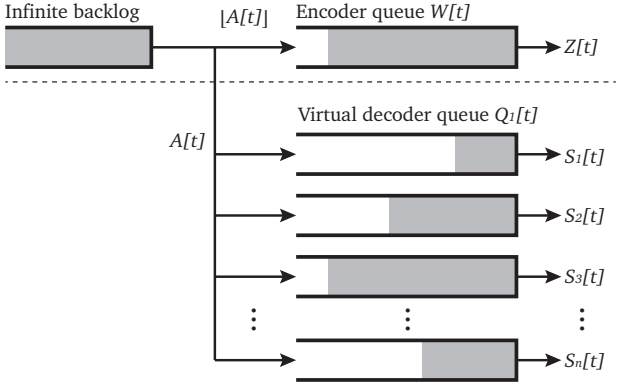}
  %\vspace{-0.25cm}
  \caption{The queueing model for MWNC-AF.}
  \vspace{-0.5cm}
  \label{fig:decode example}
\end{figure}

\begin{lemma}\label{lem_feedback}
The beacon-based anonymous feedback Algorithm \ref{feedbackalg}
satisfies
\end{lemma}
\begin{eqnarray} Z[t] = \min_{1\leq i\leq n} S_i[t]\label{Z_S1},
\end{eqnarray}
for all time-slots $t$.
\ifreport \iffinal The proof of Lemma
\ref{lem_feedback} is relegated to \cite{tech_report} due to space
limitations. \else
\begin{proof}
See Appendix~\ref{pf_lemma_0}.
\end{proof} \fi
\else The proof of Lemma \ref{lem_feedback} is relegated to
\cite{tech_report} due to space limitations. \fi

Therefore, this anonymous feedback scheme not only ensures reliable
multicast, but also keeps the encoder buffer as small as possible.

\begin{remark} In practice, the length of beacon sub-slot should
take into account the round-trip time of the beacon signal, and the
delay due to the signal detection or the hardware reaction time.
Although the beacon sub-slots are reserved in this paper, anonymous
feedback can also be implemented on a dedicated feedback channel of
orthogonal frequency.

It is important to note that the overhead of the anonymous feedback
can be significantly reduced by performing feedback only once for
every $B_{AF}$ time slots. The details of infrequent anonymous
feedback for MWNC will be discussed in Section \ref{sec:IAF}.
\end{remark}

Equations~\eqref{S_def} and~\eqref{Z_S1} tell us that
\begin{eqnarray}\label{ineq_1}
Z[t-1]\leq \min_{1\leq i\leq n} S_i[t]\leq Z[t-1] + 1.
\end{eqnarray}
Moreover, we have
\begin{eqnarray}\label{ineq_2}
\left\lfloor A[t]\right\rfloor\leq A[t]\leq \left\lfloor A[t]\right\rfloor + 1.
\end{eqnarray}
Combining Equations~\eqref{eq_W}, \eqref{decoder_queue},
\eqref{ineq_1} and~\eqref{ineq_2}, it is easy to derive
\begin{align}
\max_{1\leq i\leq n}Q_i[t]-1\leq W[t]\leq \max_{1\leq
i\leq n}Q_i[t]+1. \label{Wt_bounds}
\end{align}
The relationship between the encoding window size $W[t]$ and the
decoder queue $Q_i[t]$ is depicted in Fig. \ref{fig:decode example},
as will be clarified subsequently. One can observe that the
difference between the encoder queue length $W[t]$ and the maximum
decoder queue length $\max_{1\leq i\leq n}Q_i[t]$ is quite small.

In order to keep the queueing system stable, we assume that the
average injection rate $\mathbb{E}\{a[t]\}=\lambda$ is smaller than
the capacity, i.e., $\lambda<\inf_{1\le i\le n}\gamma_i$ for any
number of receivers $n$. We define $\gamma \triangleq \inf\{
\gamma_i, i = 1,2,\cdots\} >0$ as a lower bound of the multicast
capacity for all $n$, and $\rho \triangleq\frac{\lambda}{\gamma}$ as
the traffic intensity of the system satisfying $\rho < 1$.

\section{Performance Analysis of MWNC-AF}\label{sec:scaling law}
In this section, we rigorously analyze the decoding delay, encoding
complexity, and decoding complexity of MWNC-AF for a given
throughput $\mathbb{E}[a[t]]=\lambda~\text{packet}/\text{slot}$.
%The scaling laws of the above metrics are derived with respect to the receivers' number $n$, and are then compared with the scaling laws achieved by RLNC.

\subsection{Decoding Delay}
%As a summary of the delay analysis, we have the two theorems below.
%
%In the first place, we study the probability that a receiver
%encounters a decoding delay exceeding a given threshold in the
%asymptotic regime.

Let the time-slots $t_i^j$ $(j=1,2,\cdots)$ be the decoding moments
of receiver $i$ satisfying Equation~\eqref{def_decoding_moment}.
Suppose that packet $p_m$ is assembled at the encoder buffer in
time-slot $t$, which is between two successive decoding moments
$t_i^{j}<t\le t_i^{j+1}$. Then, packet $p_m$ will be decoded in
time-slot $t_i^{j+1}$. The decoding delay of packet $p_m$ at the
receiver $i$ is
\begin{align}
D_{i,m}=t_i^{j+1}-t.
\end{align}

Then, assuming the system is stationary and ergodic, the delay
violation probability that the decoding delay of a packet exceeds a
threshold $k$ is expressed as
\begin{align}\label{exceed_P_def}
\mathbb{P}(D_i>k)=\lim_{M\to\infty}\frac{1}{M}\sum_{m=1}^{M}{1}_{\{D_{i,m}>k\}}.
\end{align}

The average decoding delay of receiver $i$ is given by
\begin{align}\label{avD_def}
\overline{D}_i=\lim_{M\to\infty}\frac{1}{M}\sum_{m=1}^{M}D_{i,m}.
\end{align}

% for MWNC is irrelevant to the number of receivers $n$, and

%is maximized when the injection process is constant,

%The asymptotic decay rate of the probability of exceeding a given
%delay threshold is explicitly given.
\begin{theorem}\label{theorem1}
In a network with $n$ receivers, if the data injections $a[t]$ are
\emph{i.i.d.} across time-slots with an average rate
$\mathbb{E}[a[t]]=\lambda$ and $\lambda<\gamma$, then for any
receiver $i$, the asymptotic decay rate of the delay violation
probability of MWNC-AF is
%the decay rate of the probability that the decoding delay $D_i$ of an arbitrary receiver $i$ exceeding a level $k$ is derived as $k\to\infty$,
\[-\lim_{k\to\infty}\frac{1}{k}\log \mathbb{P}(D_i>k)=\Phi_i,\]
where $\log(\cdot)$ denotes natural logarithm and
\begin{align}\label{Ia_def}
\Phi_i=\sup_{\theta\in\mathbb{R}}\left\{-\log\mathbb{E}\left(e^{-\theta
a[t]}\right)-\log\left(\gamma_i e^\theta+1-\gamma_i\right)\right\}.
\end{align}
In addition,
\begin{align}
\Phi_i\le
\lambda_i\log\frac{\lambda}{\gamma_i}+(1-\lambda)\log\frac{1-\lambda}{1-\gamma_i},
\notag
\end{align}
where the equality holds if $a[t]=\lambda$ for all $t$.
\end{theorem}

\begin{proof}
See Appendix~\ref{sec:pftheorem1}.
\end{proof}

Theorem \ref{theorem1} has characterized the asymptotic decay rate
of the delay violation probability $\mathbb{P}(D_i>k)$ of receiver
$i$ as $k$ increases. It tells us that a constant packet injection
process, i.e.,
\begin{align}
a[t] =\lambda,~\forall~t,
\end{align}
achieves the fastest decay rate among all \emph{i.i.d.} packet
injection processes. We note that the decoding delay of receiver $i$
is independent of the channel condition $\gamma_j$ ($j\neq i$) of
other receivers. The reason for this is the following: By
Equation~\eqref{def_decoding_moment}, the decoding moment of
receiver $i$ is determined by $\left\lfloor Q_i[t] \right\rfloor =
0$. Further, according to Equations~\eqref{At_def},
\eqref{decoder_queue}, and \eqref{S_def}, the evolutions of $Q_i[t]$
depend on the common data injection process $A[t]$ and channel
conditions $c_i[t]$ of receiver $i$, both of which is independent of
$\gamma_j$ for $j\neq i$. In \cite{block_feedback}, the authors
derived the same expression of the delay's decay rate for the
constant injection process, which is a special case of our result.

%\begin{remark}\label{theo1_remark}
%In [??] and [??], the authors analyzed the delay performance of MWNC under a Bernoulli injection process with $a[t]\in \{0,1\}$. However, no delay decay rate result was derived. Interestingly, Theorem~\ref{theorem1} tells us that a constant injection process has a faster delay decay rate than the Bernoulli injection process. {\color{red} It would be good if we plot the two decay rates?}
%\end{remark}

%As Theorem~(\ref{theorem1} indicates, the constant injection
%process $a[t]=\lambda,\forall t$ maximizes the decay of delay
%exceeding probability. Hence, in the following, we focus on the
%delay optimal scheme.

%The following theorem characterizes the average decoding delay $\overline{D}_i$ under the constant packet injection process.
\begin{theorem}\label{theorem2} In a network with $n$
receivers, if the amount of packet injected in each time-slot is
$a[t]=\lambda$ for all $t$ and $\lambda<\gamma$, then for any
receiver $i$, the average decoding delay of MWNC-AF is upper bounded
by
\begin{align}\label{avD_det_bound}
&\overline{D}^{\text{Con}}_i\le\frac{\gamma_i(1-\gamma_i)}{2(\gamma_i-\lambda)^2}+\frac{1}{\gamma_i-\lambda}+\frac{5}{2\lambda}.
\end{align}

In addition, as $\rho$ increases to $1$,
$\overline{D}^{\text{Con}}_i$ is asymptotically upper bounded by
\begin{align}\label{avD_det_limit}
\lim_{\rho\to 1^-} \frac{\overline{D}_i^{\text{Con}}}{1/(1-\rho)^2}
\le \frac{1-\gamma}{2\gamma}.
\end{align}
\end{theorem}

\begin{proof}
See Appendix~\ref{sec:pftheorem2}.
\end{proof}

The analysis of \cite{sundararajan2009feedback} implies that, under
a Bernoulli packet injection process, i.e.,
\begin{align}
\mathbb{P}(a[t]=1) = \lambda,~~ \mathbb{P}(a[t]=0) = 1-\lambda,
\end{align}
the average decoding delay $\overline{D}_i^{\text{Ber}}$ of the
receiver $i$ with $\gamma_i=\gamma$ satisfies
\begin{align}\label{ber_limit}
\lim_{\rho\to
1^-}\frac{\overline{D}_i^{\text{Ber}}}{1/(1-\rho)^2}=\frac{1-\gamma}{\gamma}.
\end{align}
This and \eqref{avD_det_limit} tell us that for the bottleneck
receiver(s), the average decoding delay under a constant injection
process is at most one half of that of the Bernoulli packet
injection process as $\rho$ approaches $1$.

It is known that the average decoding delay of batch-based rateless
codes scales up at a speed no smaller than $O(\log n)$, as the
number of receivers $n$ increases \cite{swapna,yang2012throughput}.
Theorems \ref{theorem1} and \ref{theorem2} tell us that the decoding
delay of MWNC-AF remains constant for any number of receiver $n$.
The average decoding delay performance of two ARQ-based coding
schemes in \cite{sundararajan2009feedback,Ton13} is also bounded by
some constant independent of $n$.
%{\color{red}The difference is because, in rateless codes, no new data packets are admitted to the encoder buffer until the previous packets in the encoder buffer have been successfully delivered to all $n$ receivers. On the other hand, MWNC allows to keep injecting new data packets into the encoder buffer at a rate $\lambda$ irrelevant of $n$, such that the receivers can get access to new information.}
%The packet injection rate $\lambda$ of MWNC is irrelevant of the number of receivers $n$.
As we have mentioned, the overhead of our anonymous feedback
mechanism remains constant for any number of receiver $n$. But the
feedback overhead of the schemes in
\cite{sundararajan2009feedback,Ton13} scales up as $n$ increases.

% requires all the receiver to decoded a coded block before the next code. On the other hand, MWNC   needs to w the decoding moments of receiver $i$ is determined by only $A[t]$ and $S_i[t]$, and does not depend on the decoding state $S_j[t]$ of another receiver

%We note that it is possible to further reduce the decoding delay of MWNC. In \cite{Ton13}, the authors assumed that each receiver sends an one-bit ACK back to the transmitter in each time-slot. A smaller decoding delay is achieved by using this ACK information.
%
%
%
%We note that the our transmission scheme only requires a constant feedback overhead for any number of receivers $n$.
%
%In \cite{Ton13}, each receiver sends an one-bit ACK back to the transmitter at the end of each time slot, and this ACK information is used to reduce the decoding delay.
%
%It is possible to further reduce the decoding delay by jointly controlling the packet injection process and channel coding. For example, in \cite{Ton13}, each receiver sends an one-bit ACK back to the transmitter at the end of each time slot, and this ACK information is used to reduce the decoding delay. However, the benefit of shorter delay worse derived at the cost of unscalable feedback overhead, i.e., $O(n)$. In the following, we will show that MWNC can achieve scalable decoding delay and scalable feedback simultaneously. {\color{red} Now introduce the result on feedback overhead.}

\subsection{Encoding Complexity}
We count one operation as one time of addition and multiplication on
the Galois field. According to \eqref{rand_linear_comb} and
\eqref{eq_W}, the encoding complexity of packet $x[t]$ is ${W}[t]$,
i.e., the number of fully assembled packets in the encoder buffer.
For any given number of receivers $n$, the average encoding
complexity of MWNC-AF to encode one coded packet is
\begin{align}
\overline{W}_n=\lim_{M\to\infty}\frac{1}{M}\sum_{t=1}^{M}{W}[t],\label{def_encoding_com}
\end{align}
In addition, the probability that the encoding complexity of MWNC-AF
exceeds a threshold $k$ is depicted by
\begin{align}\label{W_exceed_P_def}
\mathbb{P}(W_n>k)=\lim_{M\to\infty}\frac{1}{M}\sum_{t=1}^{M}{1}_{\{W[t]>k\}}.
\end{align}

%We have the following theorem about encoding complexity.

\begin{theorem}\label{theorem3}
In a network with $n$ receivers, if the amount of packet injected in
each time-slot is $a[t]=\lambda$ for all $t$ and $\lambda<\gamma$,
then the average encoding complexity of MWNC-AF satisfies
%
%the average encoding complexity
%$\overline{W}_n$ of MWNC with the constant injection process
%scales as $O(\log n)$. More specifically,
\begin{equation}\label{avW_order}
\lim_{n\to\infty}\frac{\overline{W}_n}{\log n}\le\frac{1}{\eta},
\end{equation}
where %the factor $\eta$ is formally defined as {\color{red} This Eq. should look like (15)}
\begin{align}\label{eq:rate_queue}
\eta=\log \frac{\gamma e^{\theta}}{1-(1-\gamma)e^\theta},
\end{align}
and $\theta$ is the unique solution of the equation
\begin{align}\label{eq:theta_equation}
e^{-\frac{\theta}{\lambda}}\cdot\frac{\gamma
e^\theta}{1-(1-\gamma)e^\theta}=1,\quad 0<\theta<-\log (1-\gamma).
\end{align}

%{\color{red}
%Any region for $\theta$? } and
%$G_A(\theta)=\mathbb{E}\left(e^{\theta
%T_A}\right)=e^{\theta/\lambda}$ and
%$G_S(\theta)=\mathbb{E}\left(e^{\theta T_S}\right)=\frac{\gamma
%e^{\theta}}{1-(1-\gamma)e^\theta}$ are the moment generating
%functions for $T_A=1/\lambda$ and $T_S$ which is geometrically
%distributed random variable with parameter $\gamma$ respectively.

The asymptotic decay rate of the probability that the encoding
complexity exceeds a threshold is lower bounded by
\begin{align}\label{W_decay_rate}
-\lim_{k\to\infty}\frac{1}{k}\log\mathbb{P}\left(W_n>k\right)\ge\eta.
\end{align}

The inequalities in Equations~\eqref{avW_order} and
\eqref{W_decay_rate} become equalities when
$\gamma_1=\dots=\gamma_n=\gamma$.
\end{theorem}

\begin{proof}
See Appendix~\ref{sec:pftheorem3}.
\end{proof}

Theorem~\ref{theorem3} tells us that the average encoding complexity
of MWNC-AF increases as $O(\log n)$ when $n$ increases, and the
asymptotic decay rate of the encoding complexity of MWNC-AF does not
depend on $n$.

In \cite{Cogill2011}, it was shown that, for any coding scheme of
wireless multicast, the average encoder queue length must scale up
at a speed no slower than $O(\log n)$ as $n$ increases. This,
together with Theorem \ref{theorem3}, tells us that MWNC-AF has
achieved the optimal scaling law of the average encoder queue
length. Interestingly, in MWNC-AF, a large encoder queue length does
not necessarily transform into a long decoding delay, because the
encoder buffer contains both the packets that have and have not been
decoded by each receiver.

According to \cite{Cogill2011}, the encoder queue length of RLNC
also grows at a speed of $O(\log n)$.

It is worthwhile to mention that from Equation~\eqref{W_decay_rate},
the probability that the encoder queue size $W[t]$ exceeds a
threshold $k$ decays exponentially when $k$ is sufficiently large.
Therefore, the encoder queue size $W[t]$ is unlikely to be much
greater than its average value.

%{\color{red}Our simulations suggest that for relative small values of $n$, the average encoding complexity of MWNC is much smaller than that of RLNC; See Section \ref{sec:simulations} for the details.}

%More importantly, the increasing rate of the encoding complexity
%with respect to $\log n$ is explicitly given. It can be seen that as
%the injection rate $\lambda$ increases, the corresponding $\eta$
%decreases yielding a larger increasing rate of the encoding
%complexity. Reversely, $\eta$ may enable us to predict the network
%size without explicit feedback given the average encoding complexity
%at the broadcaster.
%\end{remark}

%\begin{remark}\label{theo3_remark_2}
%The second half of Theorem~\ref{theorem3} concludes that the
%probability that the encoding complexity exceeds a large enough
%threshold $k$ always drops almost exponentially with a fixed rate
%independent of the number of receivers. This is attractive, because
%it suggests that the decay rate is not affected by the number of
%receivers and in general it is unlikely to encounter encoding
%complexity much greater than the average value.
%
%Furthermore, it is found that different rate $\lambda$ of packet
%injections corresponds to different decay rate for decoding delay by
%Theorem~\ref{theorem1} and the different decay rate for encoding
%complexity by Theorem~\ref{theorem3}. Hence,
%Equations~(\ref{W_decay_rate}) and (\ref{Ia_def}) disclose a
%fundamental connection between the distribution of encoding
%complexity and that of the decoding delay.
%\end{remark}

\subsection{Decoding Complexity}
For any given number of receivers $n$, the average decoding
complexity $\Omega_n$ of MWNC-AF is measured by the average number
of operations for decoding one data packet at each receiver.

\begin{theorem}\label{theorem4}
In a network with $n$ receivers, if the amount of packets injected
in each time-slot is $a[t]=\lambda$ for all $t$ and
$\lambda<\gamma$, then the average decoding complexity of MWNC-AF,
denoted as $\overline{\Omega}_n$, satisfies
\begin{align}\label{avOmega_order}
\lim_{n\to\infty}\frac{\overline{\Omega}_n}{\log
n}\le\frac{1}{\eta},
\end{align}
where $\eta$ is defined in Equation~(\ref{eq:rate_queue}).

The inequality in Equation~\eqref{avOmega_order} becomes equalities
when $\gamma_1=\dots=\gamma_n=\gamma$.
\end{theorem}

\begin{proof}
See Appendix~\ref{sec:pftheorem4}.
\end{proof}

Theorem~\ref{theorem4} has characterized the average decoding
complexity of MWNC-AF. Interestingly, we can observe from
Equations~\eqref{avW_order} and \eqref{avOmega_order} that both the
average encoding and the average decoding complexity are of the form
$\frac{1}{\eta}\log n + o(\log n)$.

For RLNC, in order to maintain a constant throughput $\lambda>0$ as
the number of receivers $n$ increases, the average decoding
complexity of RLNC needs to increase at a rate no slower than
$O((\log n)^2)$.\footnote{The reason for this is as follows:
Consider a RLNC code with a block size of $B$ data packets. Its
average decoding complexity for each packet is of the order
$O(B^2)$, as shown in \cite{bats_code}. On the other hand, it was
shown in \cite{swapna} that in order to maintain a constant
throughput $\lambda>0$ as $n$ increases, the block size $B$ must
scale up at a speed of $O(\log n)$.} This and Theorem~\ref{theorem4}
tell us that the average decoding complexity of MWNC-AF scales much
slower than that of RLNC.

%\footnote{The reason for this is as follows: Consider a RLNC code
%with a block size of $B$ data packets. Its average decoding
%complexity for each packet is of the order $O(B^2)$, as shown in
%\cite{bats_code}. On the other hand, it was shown in \cite{swapna}
%that in order to maintain a constant throughput $\lambda>0$ as $n$
%increases, the block size $B$ must scale up at a speed of $O(\log
%n)$.}

%\subsection{Heterogeneous Packet Erasure Probabilities}\label{sec:discussion}
%We now consider the wireless multicast networks with heterogeneous
%packet erasure probabilities for different receivers. Define
%$\gamma_i\triangleq\mathbb{P}\left(c_i[t]=1\right)$ as the packet
%deliver probability of receiver $i$. By substituting $\gamma$  in Theorems~\ref{theorem1} and~\ref{theorem2} with $\gamma_i$, the delay characterization for the case of heterogeneous
%packet erasure probabilities is attained.
%
%For Theorems~\ref{theorem3} and \ref{theorem4}, we assume that the
%capacity of the multicast system has a lower bound
%$\gamma=\inf\{\gamma_i, i=1,2,\cdots \}>0$ as the number of receivers $n$ grows to infinite. This assumption is satisfied in practical multicast systems, because the receiver with a very small $\gamma_i$ cannot
%receive the multicast service reliably, and thus will disconnect from the system
%\cite{Luby2008}. By substituting $\gamma$ in Theorems~\ref{theorem3} and \ref{theorem4} with $\gamma$, we derive the upper bounds of the asymptotic encoding and decoding
%complexity performance.

\subsection{MWNC with Infrequent Anonymous Feedback}\label{sec:IAF}
So far, the anonymous feedback is performed on a per-packet basis.
Although the feedback overhead has been a constant independent of
the number of receivers, implementing feedback for every time slot
may still consume nonnegligible bandwidth resources. In this
subsection, we show that by infrequent anonymous feedback, the
feedback overhead can be conceptually reduced to $1/B_{AF}$ of that
of the original MWNC-AF, and meanwhile neither the delay nor the
reliability at the receivers is jeopardized. The costs for the
further reduction of feedback overhead are the increased encoding
and decoding complexity. The infrequent anonymous feedback provides
a tradeoff between computation complexity and feedback overhead for
MWNC-AF.

%With infrequent feedback, $B_{AF}$ packet transmissions and a single
%feedback stage comprise a frame, as shown in Figure
%\ref{fig:batched_feedback}. Same as the original MWNC-AF, packets
%are injected into the encoder according to
%$\{a[\tau]\}_{\tau\in\mathbb{N}}$. The main difference is that in
%the anonymous feedback stage for a frame, a receiver $i$ would send
%a beacon signal if and only if $S_i[t]=Z_i[t-1]+B_{AF}-1$, where $t$
%is the last time slot in the current frame. Once beacon signal is
%detected at the transmitter, no packets would be removed from the
%encoder in all time slots of the next frame, i.e., $Z[\tau]=Z[t-1]$
%for $t\le\tau\le t+B_{AF}-1$. If no beacon signal is detected at the
%transmitter, a block of $B_{AF}$ packets would be removed from the
%encoder in the first slot of the next frame, i.e.,
%$Z[\tau]=Z[t-1]+B_{AF}$ for $t\le\tau\le t+B_{AF}-1$. The intuition
%behind is simple. In the event that $S_i[t]>Z_i[t-1]+B_{AF}-1$, even
%if a block of $B_{AF}$ packets is removed and no new packets are
%received in the next frame, Equation~\eqref{Z_S0} can be still
%satisfied in the next frame and reliable multicast can be ensured.

\begin{figure}[t]
\centering
  \includegraphics[width=3.5in]{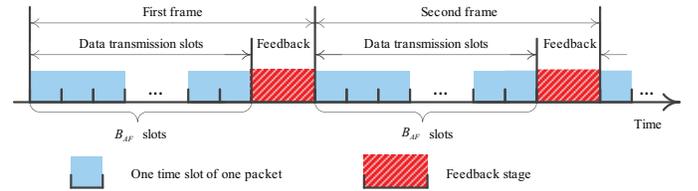}
  %\vspace{-0.25cm}
  \caption{Illustration of MWNC with Infrequent Anonymous Feedback.}
  \vspace{-0.5cm}
  \label{fig:batched_feedback}
\end{figure}

In this policy, anonymous feedback is practiced once for a frame of
$B_{AF}$ packet transmissions, as shown in Figure
\ref{fig:batched_feedback}. If the transmitter cannot detect the
beacon signal, it will remove $B_{AF}$ packets from the encoder
buffer at the end of the frame. Otherwise, no packet will be
removed. We can ensure that the removed packets are already ``seen''
at each receiver, i.e., the multicast transmissions are reliable. We
note that this policy does not increase the decoding delay, because
the decoding delay is determined by the virtual decoder queue
$Q_i[t]$ of each receiver, and does not depend on the encoder queue
$W[t]$.

Due to the infrequent removal of packets in the encoder, the average
encoding as well as decoding complexity of MWNC-AF with $B_{AF}>1$
would be greater than the case when $B_{AF}=1$. However, it is
straightforward to see that with infrequent anonymous feedback, the
average encoding complexity is at most $B_{AF}$ more than the
average encoding complexity of the original MWNC-AF, i.e.,
$\overline{W}_n+B_{AF}$.

%Besides, the feedback channel has been assumed to be reliable, i.e.,
%the base station can reliably detect the beacon signal from the
%receivers and a receiver can reliably detect a beacon signal sent by
%the base station. In practice, although the control signals can be
%sent in a reliable manner, i.e., low data rate or additional error
%correction codes, the overhead to guarantee high reliability for the
%feedback signaling could be nontrivial.

\section{Numerical Results}\label{sec:simulations}
This section presents some simulation results that provide insights
and trends as well as validate the theoretical results. We
investigate three important aspects of performance: decoding delay,
encoding complexity, and decoding complexity. We consider two
network scenarios, one with homogeneous channel conditions where
$\gamma_1=\dots=\gamma_n=0.6$, and the other with heterogenous
channel conditions where $\gamma_1=\dots=\gamma_{n/2}=0.6$ and
$\gamma_{n/2+1}=\dots=\gamma_{n}=0.8$. The simulation results are
derived by running over at least $10^7$ time-slots.

% you provide numerical figures that provide insights and trends as well as validate the theoretical results.

\begin{figure}[t]
\centering
  \includegraphics[width=3.5in]{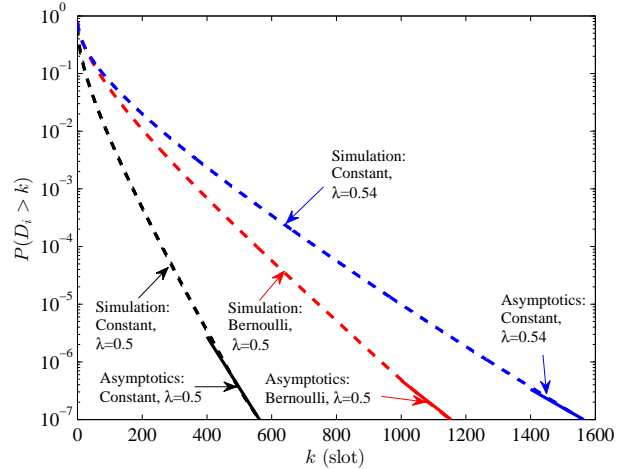}
  %\vspace{-0.25cm}
  \caption{Simulation results of the delay violation probability $\mathbb{P}(D_i>k)$ of MWNC-AF versus
$k$ for $\gamma_i=0.6$.}
  %\vspace{-0.1cm}
  \label{fig:delay_decay}
\end{figure}

\begin{figure}[t]
\centering
  \includegraphics[width=3.5in]{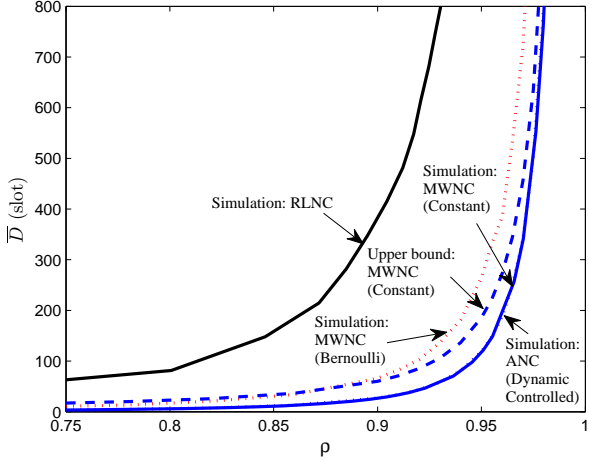}
  %\vspace{-0.25cm}
  \caption{Simulation results of the average decoding delay $\overline{D}_i$ versus
the traffic intensity $\rho$ for $n=100$ and
$\gamma_1=\dots=\gamma_n=0.6$. The average decoding delay of
ARQ-based network coding (ANC) with dynamic injection control
\cite{Ton13} is very close to MWNC-AF. However, its feedback
overhead grows linearly with $n$, while our scheme only requires a
fixed amount of feedback overhead.
%with constant injection process achieves a throughput-delay tradeoff
%very close to that of the dynamically controlled method in
%, yet incurs only $O(1)$ feedback overhead in contrast
%to the $O(n)$ feedback overhead which is necessary in the
%dynamically controlled method.
}
  %\vspace{-0.1cm}
  \label{fig:TD_tradeoff}
\end{figure}

\begin{figure}[t]
\centering
  \includegraphics[width=3.5in]{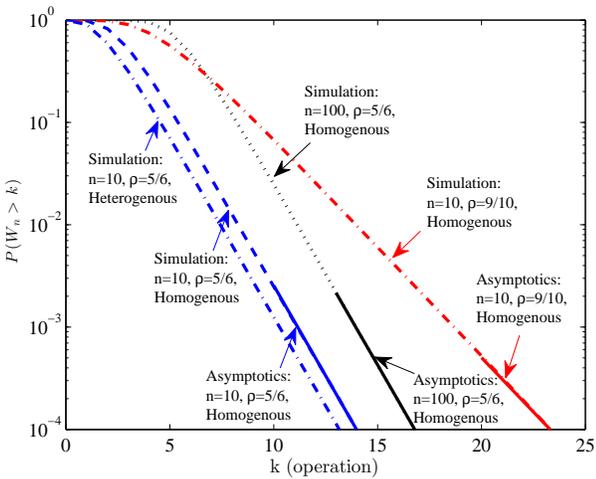}
  %\vspace{-0.25cm}
  \caption{Simulation results of the buffer overflow probability $\mathbb{P}(W_n>k)$ of MWNC-AF (constant injection process) versus
$k$ for $\gamma=0.6$.}
  %\vspace{-0.1cm}
  \label{fig:W_decay}
\end{figure}

\begin{figure}[t]
\centering
  \includegraphics[width=3.5in]{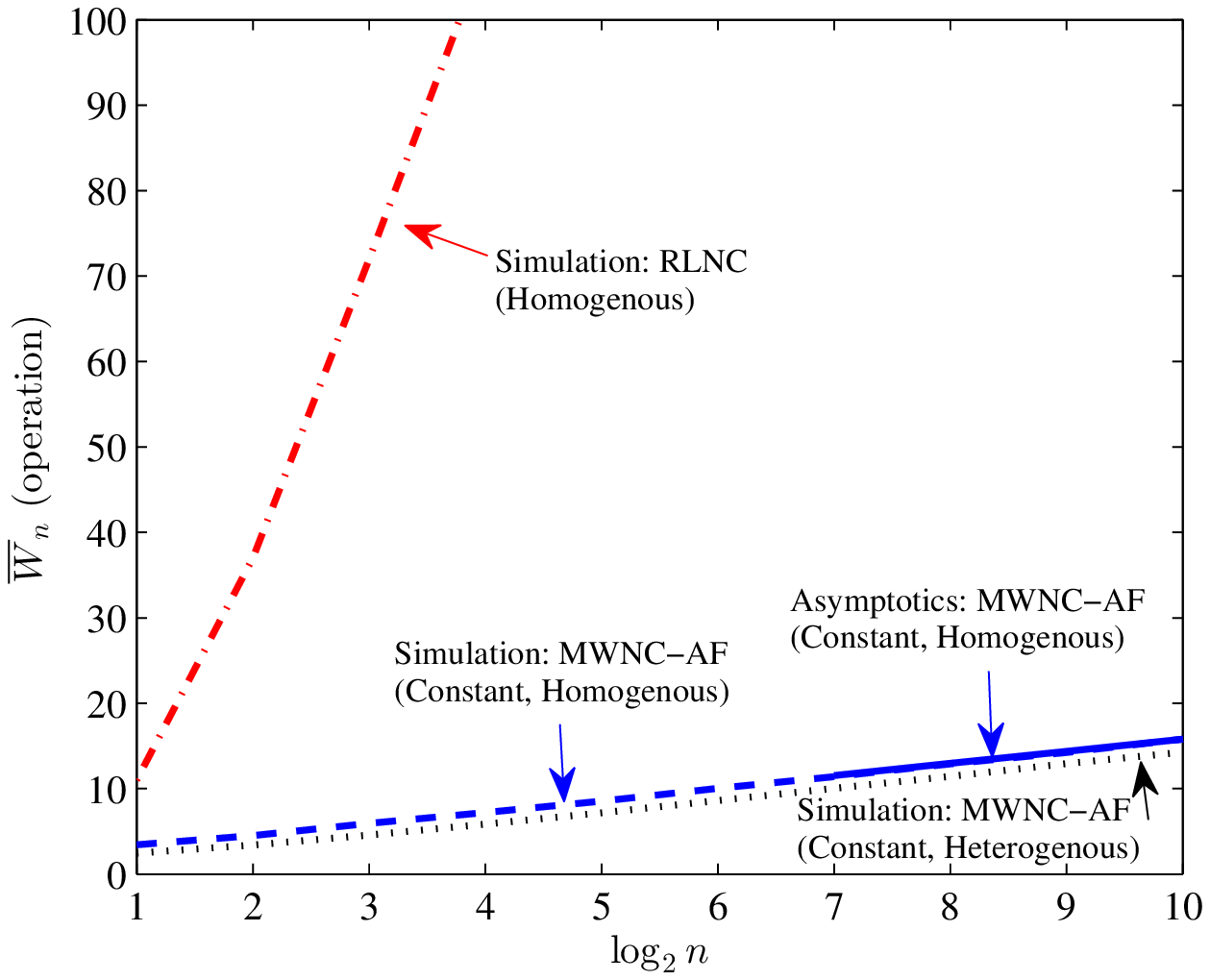}
  %\vspace{-0.25cm}
  \caption{Simulation results of the average encoding complexity $\overline{W}_n$ versus the number of receivers $n$ for $\rho=0.9$ and $\gamma=0.6$.}
  %\vspace{-0.1cm}
  \label{fig:encoding_complexity}
\end{figure}

\begin{figure}[t]
\centering
  \includegraphics[width=3.5in]{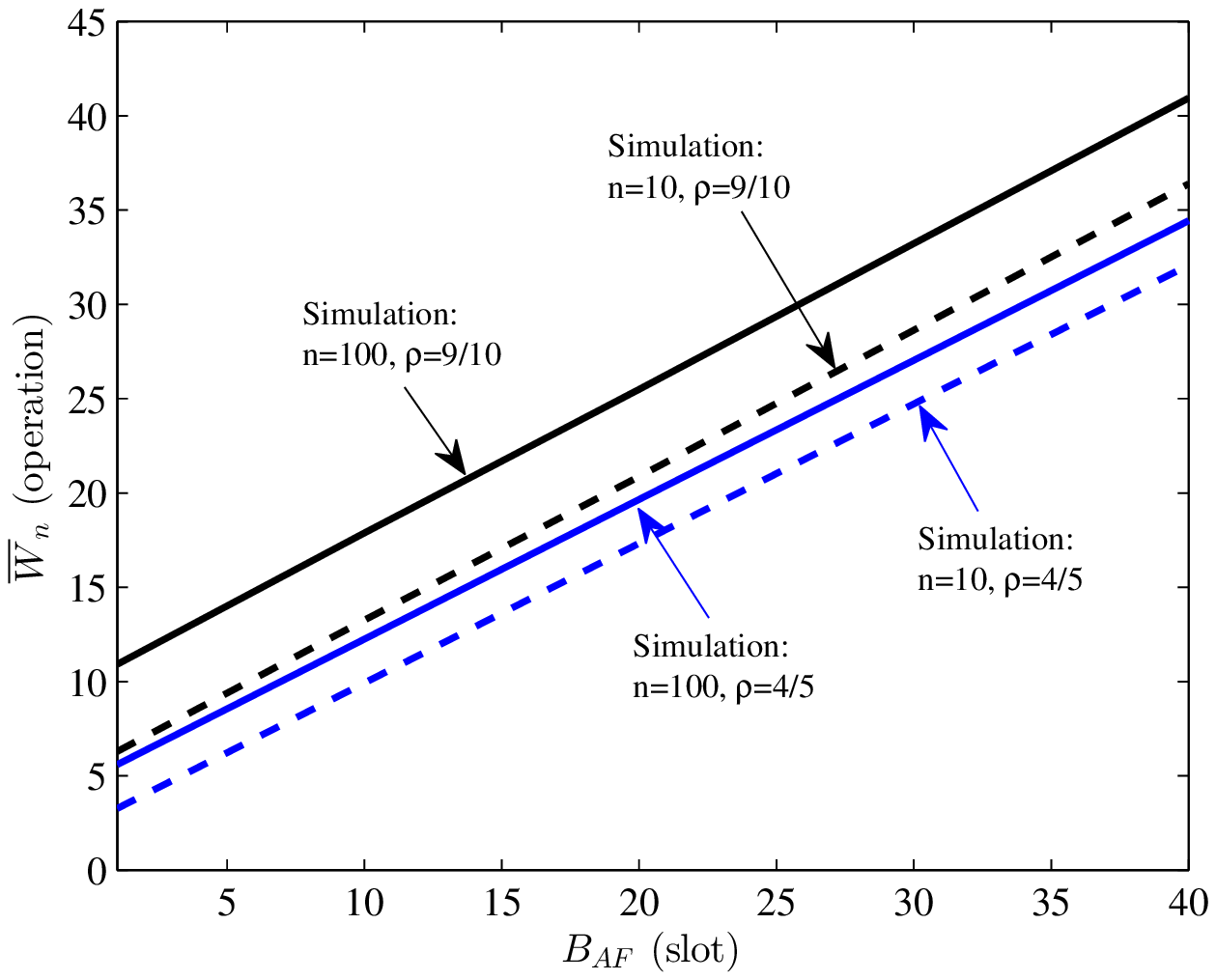}
  %\vspace{-0.25cm}
  \caption{Simulation results of the average encoding complexity $\overline{W}_n$ versus $B_{AF}$ for $\gamma_1=\dots=\gamma_n=0.6$.}
  %\vspace{-0.1cm}
  \label{fig:BAF_encoding}
\end{figure}

\begin{figure}[t]
\centering
  \includegraphics[width=3.5in]{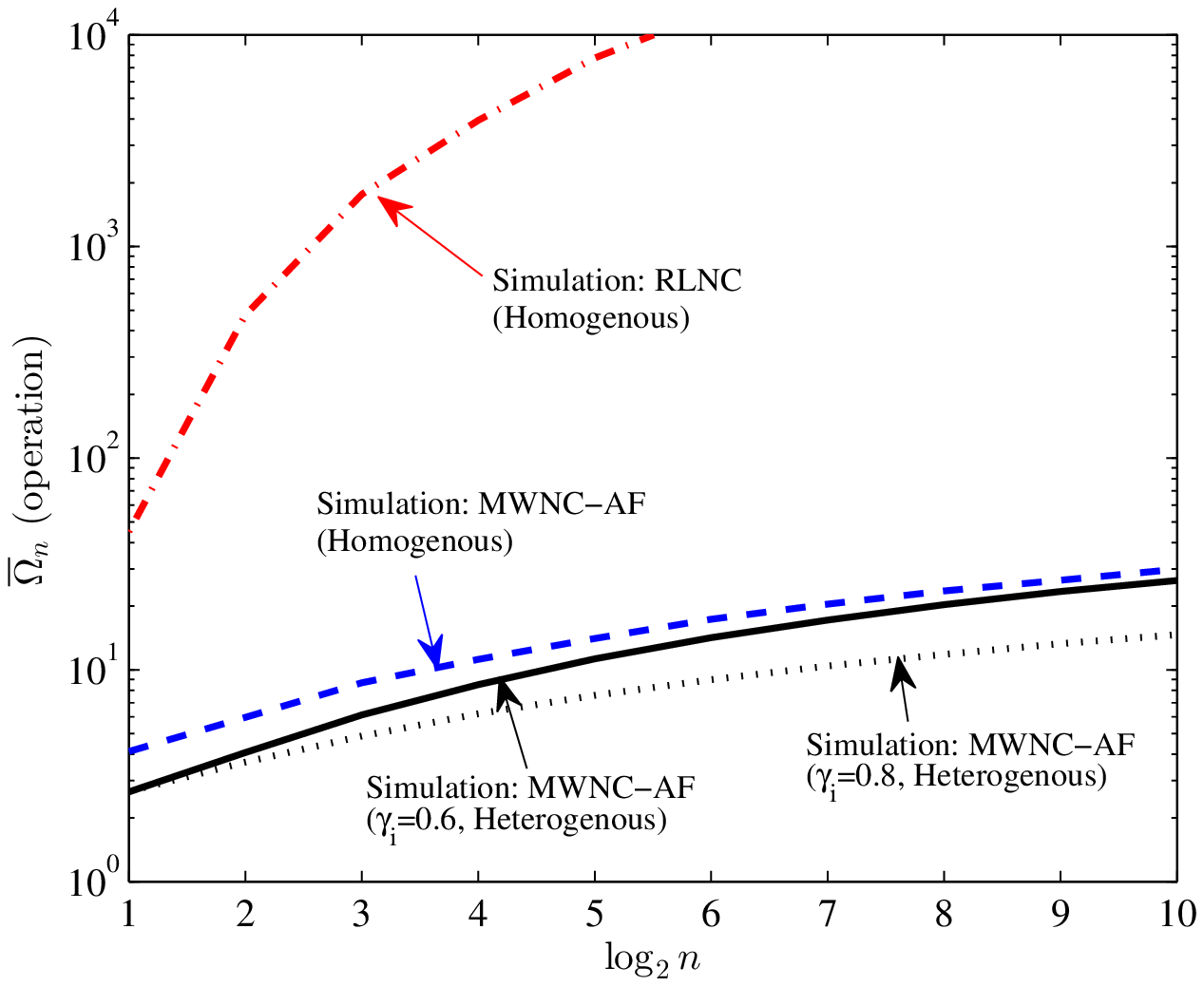}
  %\vspace{-0.25cm}
  \caption{Simulation results of the average decoding complexity $\overline{\Omega}_n$ versus the number of receivers $n$ for $\rho=0.9$ and $\gamma=0.6$.}
  %\vspace{-0.1cm}
  \label{fig:decoding_complexity}
\end{figure}

\begin{figure}[t]
\centering
  \includegraphics[width=3.5in]{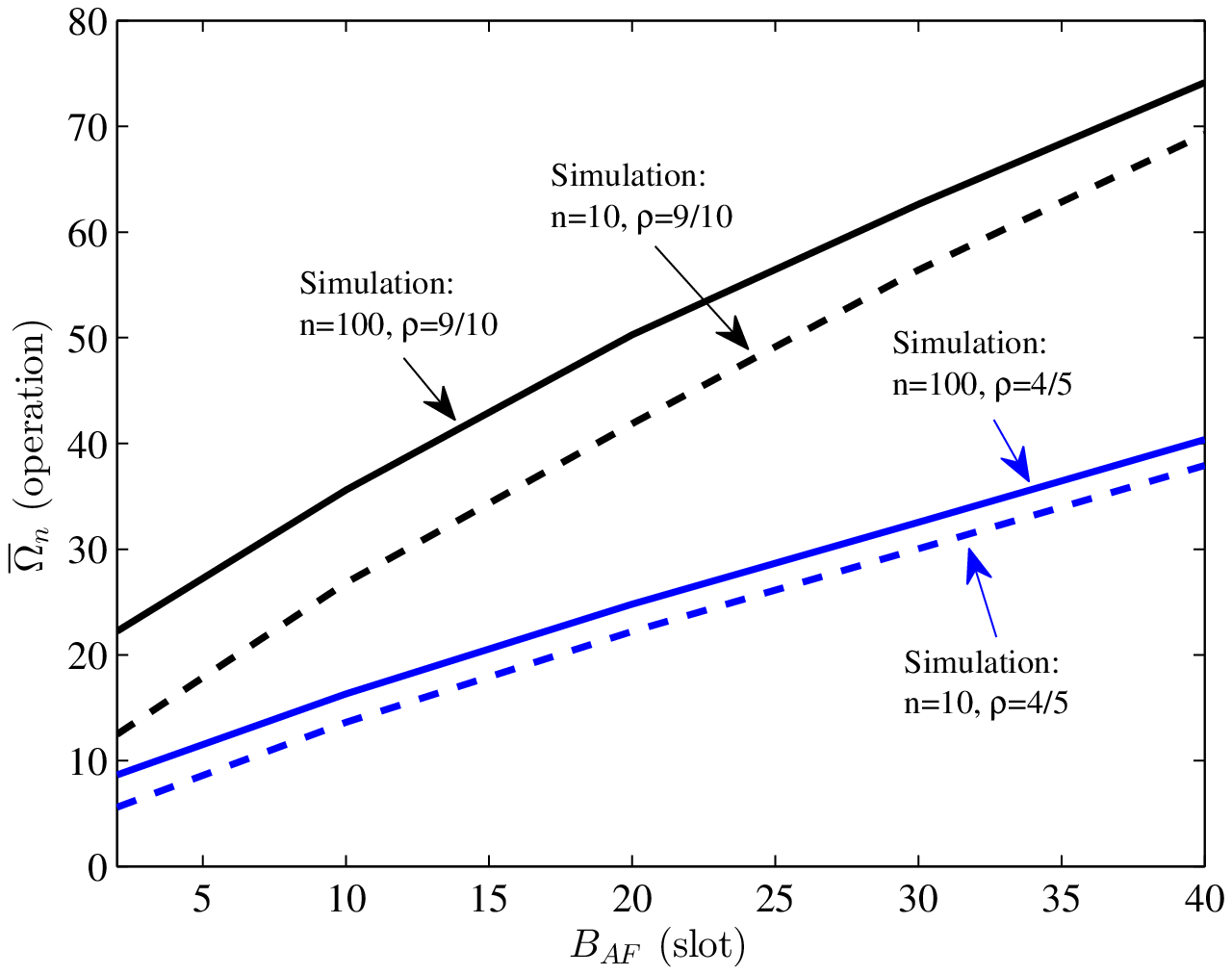}
  %\vspace{-0.25cm}
  \caption{Simulation results of the average decoding complexity $\overline{W}_n$ versus $B_{AF}$ for $\gamma_1=\dots=\gamma_n=0.6$.}
  %\vspace{-0.1cm}
  \label{fig:BAF_decoding}
\end{figure}

\subsection{Decoding delay}
Since the delay performance for a receiver of MWNC-AF is uniquely
determined by the injection process and the channel conditions of
the receiver, we focus on a receiver with $\gamma_i=0.6$. Figure
\ref{fig:delay_decay} illustrates the delay violation probability
$\mathbb{P}(D_i>k)$ of MWNC-AF versus $k$. One can observe that the
delay violation probability $\mathbb{P}(D_i>k)$ of MWNC-AF decays
exponentially for sufficiently large $k$ and matches the predicted
asymptotic decay rate from Equation~\eqref{Ia_def}. For
$\lambda=0.5$, as expected from our theoretical results, we find
that a constant packet injection process achieves a much faster
decay rate than the Bernoulli packet injection process. In addition,
comparing the simulation results for $\lambda=0.5$ and
$\lambda=0.54$, the delay violation probability $\mathbb{P}(D_i>k)$
for a fixed $k$ increases with respect to $\lambda$. Therefore,
there is a tradeoff between system throughput and delay violation
probability. One can utilize Equation~\eqref{Ia_def} to search for
the parameters $\lambda$ and $\gamma$ for achieving an appropriate
delay-throughput tradeoff depending on design requirements.

Figure \ref{fig:TD_tradeoff} plots the average decoding delay
$\overline{D}_i$ of different network coding schemes versus the
traffic intensity $\rho$ in the homogeneous network setting, where
$n=100$ and $\gamma=0.6$. One can observe the following results:
First, the average decoding delay of RLNC \cite{ho2006random} with
batched packet arrivals is much larger than that of MWNC-AF. We note
that the average decoding delay of LT codes \cite{luby2002}, Raptor
codes \cite{shokrollahi2006raptor} are larger than that of RLNC,
because of an extra reception overhead. Second, the average decoding
delay of MWNC-AF with constant packet injections is much smaller
than that of MWNC-AF with Bernoulli packet injections. When $\rho$
tends to 1, the constant packet injection process can reduce the
average decoding delay of MWNC-AF by one half, over the Bernoulli
packet injection process. Third, the average decoding delay of
ARQ-based network coding (ANC) with dynamic injection control
\cite{Ton13} is almost the same as that of MWNC-AF with constant
packet injections. However, it is important to note that the scheme
of \cite{Ton13} requires explicit feedback from each receiver, and
thus its total feedback overhead grows as $O(n)$. In comparison, the
feedback overhead of MWNC-AF with constant packet injections remains
the same, regardless of $n$. Finally, the delay upper bound in
Equation~\eqref{avD_det_bound} for MWNC-AF with constant packet
injections is accurate for high load.

% the tradeoff between
%throughput and average decoding delay of RLNC \cite{ho2006random},
%ANC \cite{kumar2008arq}, the Dynamic Controlled scheme \cite{Ton13}
%and MWNC. The number of receivers is set to be $n=100$. Firstly, it
%can be seen that given the same throughput, the average decoding
%delay of RLNC is much larger than that of the other schemes, which
%shows the limitation of the batch-based method in term of delay.
%Secondly, among the online coding schemes, the performance of MWNC
%with deterministic injections is found to be almost the same as the
%Dynamic Controlled method in \cite{Ton13}, which achieves the best
%decoding delay performance. However, unlike \cite{Ton13} which
%requires the explicit feedback from each receiver, the delay gain of
%MWNC is achieved at $O(1)$ feedback overhead. Furthermore, it can be
%observed that the upper bound of decoding delay derived in Theorem
%\ref{theorem2} is able to provide tractable delay performance for
%MWNC. By comparing with ANC \cite{kumar2008arq} with Bernoulli
%injections, it is obvious that MWNC with deterministic injections
%can achieve about half the decoding delay of ANC under the same
%throughput, which coincides with the implication of Theorem
%\ref{theorem2}.

\subsection{Encoding complexity}

Figure \ref{fig:W_decay} plots the probability $\mathbb{P}(W_n>k)$
of MWNC-AF versus $k$ for $\gamma=0.6$. One can observe that
$\mathbb{P}(W_n>k)$ decays exponentially for sufficiently large $k$
and matches the predicted asymptotic decay rate $\eta$ from
Equation~\eqref{eq:rate_queue}. Since $\eta$ is a decreasing
function of $\rho$ and is irrelevant of $n$, the traffic intensity
$\rho$ has a larger impact on the probability $\mathbb{P}(W_n>k)$
than the number of receivers $n$, when $k$ is sufficiently large. It
can be also found that the decay rate of $\mathbb{P}(W_n>k)$ with
the heterogeneous channel conditions is very close to that with the
homogenous channel conditions.

In Fig. \ref{fig:encoding_complexity}, we compare the average
encoding complexity $\overline{W}_n$ of different network coding
schemes versus the number of receivers $n$, where $\rho=0.9$ and
$\gamma=0.6$. In the homogeneous network scenario, we find that the
increasing rate of the average encoding complexity of MWNC-AF
matches well with the predicted asymptotic rate even for relatively
small $n$. The expression $\frac{1}{\eta} \log n$ provides a close
approximation of the average encoding complexity of MWNC-AF. One can
also observe that the average encoding complexity of RLNC is of the
order $O(\log n)$, but its pre-factor is larger than that of
MWNC-AF, i.e., $1/\eta$. Therefore, the average encoding complexity
of RLNC grows faster than that of MWNC-AF as $n$ increases. When
$n=1024$ receivers, the average encoding complexity of MWNC-AF is
less than 25. In the heterogenous network scenario, the average
encoding complexity is less than but close to that in the homogenous
network setting.

In Figure \ref{fig:BAF_encoding}, we show  the impact of infrequent
anonymous feedback on the encoding complexity of MWNC-AF. It can be
seen that the average encoding complexity increases almost linearly
with respect to $B_{AF}$. Even for 100 receivers with a load as high
as 0.9, feedback can be performed only once in every 40 slots, at
the same time less than 45 operations are needed on average to
encode a packet.

\subsection{Decoding complexity}
In Fig. \ref{fig:decoding_complexity}, we compare the average
decoding complexity $\overline{\Omega}_n$ of different network
coding schemes with the number of receivers $n$ for $\rho=0.9$ and
$\gamma=0.6$. In the homogeneous network scenario, one can observe
that the average decoding complexity of RLNC is much larger than
that of MWNC-AF. Our simulation results suggest that the average
decoding complexity of MWNC-AF grows as $O(\log n)$. In particular,
as $n$ grows from 2 to 1024, the average decoding complexity of
MWNC-AF is only increased by 8 times. However, the pre-factor of the
average decoding complexity of MWNC-AF has not converged to $1/\eta$
as $n$ grows to 1024. We believe that this convergence would occur
at very large values of $n$, which is beyond our current simulation
capability. In the heterogenous network scenario, the average
decoding complexity is less than that in the homogenous network
setting.

Note that we have chosen a relative large value of $\rho$ (i.e.,
$\rho=0.9$) in Figs. \ref{fig:encoding_complexity} and
\ref{fig:decoding_complexity}. The average encoding and decoding
complexity of MWNC-AF will be even smaller as $\rho$ decreases.

Lastly, in Figure \ref{fig:BAF_decoding}, we show the impact of
infrequent anonymous feedback on the decoding complexity of MWNC-AF.
For a given $B_{AF}$ and $n$, the average decoding complexity is
larger than the average encoding complexity shown in Figure
\ref{fig:BAF_encoding}, and the difference is more evident for high
load. Even for 100 receivers with a load as high as 0.9, feedback
can be performed only once in every 40 slots, at the same time less
than 80 operations are needed on average to decode a packet.

%We also plot the curve  $\overline{W}_n=\eta \log n$, where $\eta$ is determined from Equation~\eqref{eq:rate_queue}. We find that this curve provides an accurate approximate of the average encoding complexity for MWNC.
%
%Lastly, under a fixed target target load $\rho=5/6$, we compare the
%required decoding complexity of RLNC and MWNC as the number of
%receivers $n$ increases. It can be found that the decoding
%complexity of RLNC rockets as $n$ increases which demands for larger
%batch size $B$, while the decoding complexity for MWNC increases
%almost at a moderate linear rate with respect to $\log n$. Besides,
%the slope of increase is close as the slope of increase of the
%encoding complexity. The simulation results support
%Theorem~\ref{theorem4}.

\section{Conclusions}\label{sec:conclusion}
In this paper, we have developed a joint coding and feedback scheme
called Moving Window Network Coding with Anonymous Feedback
(MWNC-AF). We have rigorously characterized the decoding delay,
encoding complexity, and decoding complexity of MWNC-AF. Our
analysis has shown that MWNC-AF achieves constant decoding delay and
constant feedback overhead for any number of receivers $n$, without
sacrificing the throughput and reliability of wireless multicast. In
addition, we have proven that injecting a fixed amount of
information bits into the MWNC-AF encoder buffer in each time-slot
can achieve a much shorter decoding delay than the Bernoulli data
injection process. We have also demonstrated that the encoding and
decoding complexity of MWNC-AF grow as $O(\log n)$ as $n$ increases.
Our simulations show that, for $n=1024$ receivers, the encoding and
decoding complexity of MWNC-AF are still quite small. Therefore,
MWNC-AF is suitable for wireless multicast with a large number of
receivers.

%In this paper, we have developed a MWNC transmission scheme with
%anonymous feedback. We have rigorously characterized the decoding
%delay, encoding complexity, and decoding complexity of the MWNC
%transmission scheme. Our analysis has shown that this MWNC
%transmission scheme achieves constant decoding delay and constant
%feedback overhead for any number of receivers $n$, without
%sacrificing the throughput and reliability of wireless multicast. In
%addition, we have proven that injecting a fixed amount of
%information bits into the MWNC encoder buffer in each time-slot can
%achieve a much shorter decoding delay than the Bernoulli data
%injection process. We have also demonstrated that the encoding and
%decoding complexity of MWNC grow as $O(\log n)$ as $n$ increases.
%Our simulations show that, for $n=1024$ receivers, the encoding and
%decoding complexity of MWNC are still quite small. Therefore, the
%MWNC transmission scheme with anonymous feedback is suitable for
%wireless multicast with a large number of receivers.

\appendices
%\section{Proofs}\label{sec:proof}
%Before elaborating the details of the proofs, we establish the dynamics of the decoder queues and introduce some common notations.

\section{preliminaries}\label{sec:proof}
%We now provide some preliminary results, which are helpful for our proofs.

We first provide some preliminary results, which are helpful for our proofs.
%show that the dynamics of the decoder queue
%$Q_i[t]$ for receiver $i$ conforms to a random walk with one
%reflecting barrier on $[0,\infty)$.

According to Equations~(\ref{At_def}) and~(\ref{decoder_queue}), we can derive
\begin{align}
Q_i[t-1]+a[t]=A[t]-S_i[t-1].
\end{align}
Using this and~(\ref{S_def}), one can derive the evolutions of the decoder queue $Q_i[t]$, given by
\begin{align}
Q_i[t]=Q_i[t-1]+a[t]-c_i[t]1_{\left\{Q_i[t-1]+a[t]\ge
1\right\}}.\label{rw_normal}
\end{align}
Accordingly, $\{Q_i[t]\}_t$ is a random walk on $[0,\infty)$, which
has a steady state distribution if $\lambda<\gamma_i$.

%When $Q_i[t-1]\ge 1$, from Equations~(\ref{At_def}), (\ref{S_def}),
%and (\ref{decoder_queue}),
%\begin{align}
%Q_i[t]=Q_i[t-1]+a[t]-C_i[t].\label{rw_normal}
%\end{align}
%Therefore, $Q_i[t]$ evolves as a random walk with a step size
%$\chi[t]=a[t]-C_i[t]$ when $Q_i[t-1]\ge 1$.
%
%When $Q_i[t-1]<1$, the evolution of $Q_i[t]$ depends on $a[t]$. By
%Equations~(\ref{At_def}), (\ref{S_def}), and (\ref{decoder_queue}),
%if
%\begin{align}
%Q_i[t-1]+a[t]<1,
%\end{align}
%we have
%\begin{align}
%Q_i[t]=Q_i[t-1]+a[t],\label{rw_case1}
%\end{align}
%$Q_i[t]$ is increased by $a[t]$. Otherwise $Q_i[t-1]+a[t]\ge 1$,
%\begin{align}
%Q_i[t]=Q_i[t-1]+a[t]-C_i[t],\label{rw_case2}
%\end{align}
%$Q_i[t]$ continues to move from $Q_i[t-1]+a[t]$ by $-C_i[t]$. Hence,
%$1$ is a reflecting barrier for $Q_i[t]$.

\begin{statement}\label{independency}
If the injection process is constant, i.e., $a[t]=\lambda$ for all $t$, then the decoder queues $\left\{Q_i[t]\right\}_{1\le i\le n}$ are
independent.
\end{statement}
When $a[t]=\lambda$ for all $t$, the injection and departure
processes $\{a[t],c_i[t]\}_t$ are independent for different
receivers. Then, Statement  \ref{independency} follows from the
queue evolution in Equation~\eqref{rw_normal}. For general packet
injection processes, the decoder queues $\left\{Q_i[t]\right\}_{1\le
i\le n}$ are correlated.

%If $a[t]=\lambda$ for all $t$, then the arrival and departure processes $\{a[t],C_i[t]\}_t$ are independent for different receivers. Therefore, the decoder queues $\left\{Q_i[t]\right\}_{1\le i\le n}$ are also independent.
%
%
%The following statement can be seen from
%Equations~(\ref{rw_normal}), (\ref{rw_case1}) and (\ref{rw_case2})

Next, we show that the decoding procedure for any receiver $i$ can be captured by a Markov renewal process. Since the system is symmetric, we only need to consider the decoding procedure at receiver 1. Let us define $T_j\triangleq t_1^{j+1}-t_1^j$. Since $\{t_1^j\}_j$ is set of the decoding moments of receiver $1$ that satisfies Equation~(\ref{def_decoding_moment}), we know that $T_j$ represents the interval between the $j^\text{th}$ decoding moment and the $(j+1)^\text{th}$ decoding moment and can be expressed as
\begin{align}
T_j=\min \{t\geq 1: Q_1[t_1^j+t]<1\}.\label{T_power_def1}
\end{align}
The value of $T_j$ depends on the queue length $Q_1[t_1^j]$ at the
$j^\text{th}$ decoding moment, which, according to the definition of
decoding moments in Equation~(\ref{def_decoding_moment}), is a value
between $0$ and $1$. By combining the above equation with
Equation~(\ref{rw_normal}), we can further rewrite the expression
for $T_j$ as
\begin{align}
T_j=\min\left\{t\geq 1: Q_1[t_1^j]+\sum_{\tau=1}^t\left(a[\tau]-c_1[\tau]\right)<1\right\},\label{T_power_def}
\end{align}
with the following reasoning: 1) If $Q_1[t_1^j]+a[t_1^j+1]\geq 1$,
then according to Equation~(\ref{rw_normal}), we know that
$Q_1[t_1^j+t]=Q_1[t_1^j]+\sum_{\tau=1}^{t}\left(a[t_1^j+\tau]-c_1[t_1^j+\tau]\right)$
as long as $Q_1[t_1^j+\tau]\geq1$ for all $\tau$ from $1$ to $t-1$.
2) If $Q_1[t_1^j]+a[t_1^j+1]<1$, then although $Q_1[t_1^j+1]\neq
Q_1[t_1^j]+a[t_1^j+1]-c[t_1^j+1]$, both $Q_1[t_1^j+1]$ and
$Q_1[t_1^j]+a[t_1^j+1]-c[t_1^j+1]$ is less than 1, Thus
Equation~\eqref{T_power_def} gives an alternative expression for
$T_j$ defined in Equation~\eqref{T_power_def1}.

Based on Equation~(\ref{T_power_def}), we can easily verify that the following equation holds:
\begin{align}
&\mathbb{P}\left(Q_1[t_1^{j+1}]\le x, T_j\le
t\Big|Q_1[t_1^{1}],...,Q_1[t_1^{j}];T_1,...,T_{j-1}\right)\nonumber\\
=&\mathbb{P}\left(Q_1[t_1^{j+1}]\le x, T_j\le
t\Big|Q_1[t_1^{j}]\right),\forall x\in[0,1),\forall
t\in\mathbb{N}.\nonumber
\end{align}
The above equation indicates that the process $\{Q_1[t_1^j],T_j\}_j$ is a Markov renewal process, where $Q_1[t_1^j]$ is the initial state of the $j^\text{th}$ renewal. Let $K_j$ denote the number of packets that are injected to the encoder queue between time-slot $t_1^j$ and time-slot $t_1^{j+1}$, then it can be expressed as
\begin{align}\label{K_j_def}
K_j=\left\lfloor Q_1[t_1^j]+
\sum_{t=t_1^{j}+1}^{t_1^{j}+T_j}a[t]\right\rfloor.
\end{align}
To facilitate the analysis of the Markov renewal process $\{Q_1[t_1^j],T_j\}_j$, we denote $\widehat{Q}_1$ as a random variable that has the same distribution as the steady state distribution of the initial state of the Markov renewal process. More precisely, $\mathbb{P}(\widehat{Q}_1>q)=\mathbb{P}(Q_1[t_1^\infty]>q)$ for any $q$.

For each $0\leq q<1$, we also define a random variable $\widehat{T}(q)$, which can be expressed as
\begin{align}
\widehat{T}(q)=\left\{t\geq
1:q+\sum_{\tau=1}^{t}(\widehat{a}[\tau]-\widehat{c}_1[\tau])<1\right\},\label{T_hat}
\end{align}
where $\{\widehat{a}[\tau]\}_\tau$ and
$\{\widehat{c}_1[\tau]\}_\tau$ are two groups of i.i.d. random
variables that have the same distributions as $a[1]$ and $c_1[1]$,
respectively. By comparing Equation~(\ref{T_hat}) with
Equation~(\ref{T_power_def}), we know that $\widehat{T}(q)$ has the
same distribution as $T_j$ when $Q_1[t_1^j]=q$. Similarly, we define
$\widehat{K}(q)\triangleq\left\lfloor q+
\sum_{t=1}^{\widehat{T}(q)}\widehat{a}[t]\right\rfloor$.

The reason why we define $\widehat{Q}_1$, $\widehat{T}(q)$, and
$\widehat{K}(q)$ will become clear later in the proofs where the
Markov renewal reward theory (Theorem 11.4 in
\cite{ccinlar1975exceptional}) is invoked. By the property of
conditional expectation, we have
\begin{align}
\mathbb{E}{\big[}\widehat{T}{\big]}&\triangleq \mathbb{E}{\Big[}\mathbb{E}{\big[}\widehat{T}(\widehat{Q}_1)|\widehat{Q}_1{\big]}{\Big]}, \notag\\
\mathbb{E}{\big[}\widehat{T}^2{\big]}&\triangleq \mathbb{E}{\Big[}\mathbb{E}{\big[}\widehat{T}(\widehat{Q}_1)^2|\widehat{Q}_1{\big]}{\Big]}, \notag\\
\mathbb{P}{\big(}\widehat{T}>k{\big)}&\triangleq \mathbb{E}{\Big[}\mathbb{P}{\big(}\widehat{T}(\widehat{Q}_1)>k|\widehat{Q}_1{\big)}{\Big]}.\label{PT_hat}
\end{align}

\section{Proof of Theorem 1}\label{sec:pftheorem1}
In this subsection, we analyze the probability that the decoding
delay experienced by a receiver exceeds a given threshold for the
coding scheme with general \emph{i.i.d.} injection processes.
Without loss of generality, we focus on the analysis of the decoding
delay of receiver $1$.

\begin{lemma}\label{lemma_D1k}
$\mathbb{P}(D_1>k)$ is upper and lower bounded by
\begin{align}\label{exceed_P_bounds}
&\frac{\mathbb{P}\left(\widehat{T}>k\right)}{\lambda\mathbb{E}{\big[}\widehat{T}{\big]}}\le\mathbb{P}(D_1>k)\le\frac{k\mathbb{P}\left(\widehat{T}>k\right)+\sum_{b=k}^\infty
\mathbb{P}\left(\widehat{T}>b\right)}{\lambda\mathbb{E}{\big[}\widehat{T}{\big]}}.
\end{align}
\end{lemma}
\begin{remark} The proof of Lemma \ref{lemma_D1k} is
based on a simple observation. For a given delay threshold $k>0$,
the number of packets decoded after an interval $T_j$ must satisfy
the following conditions. 1) If $ {T}_j\le k$, there is no packets
exceeding the threshold $k$. 2) If $ {T}_j> k$, there are at most $
{T}_j$ packets which exceed the threshold $k$. 3) If $ {T}_j> k$,
there is at least one packet which exceed the threshold $k$.
\end{remark}
\ifreport \iffinal The proof of Lemma \ref{lemma_D1k} is relegated
to \cite{tech_report} due to space limitations. \else
\begin{proof}
See Appendix~\ref{pf_lemma_1}.
\end{proof} \fi
\else The proof of Lemma \ref{lemma_D1k} is relegated to
\cite{tech_report} due to space limitations. \fi

Lemma \ref{lemma_D1k} shows the connection between $\mathbb{P}(D_1>k)$
and $\mathbb{P}\left(\widehat{T}>b\right)$. Hence, subsequently we
study the probability that the decoding interval in the steady state
exceeds a certain threshold, i.e.,
$\mathbb{P}\left(\widehat{T}>b\right),b\in\mathbb{N}$.

\begin{lemma}\label{lemma_Ia}
The decay rate of the decoding interval in the steady state is given
by
\begin{align}\label{interval_rate_final}
-\lim_{b\to\infty}\frac{1}{b}\log
\mathbb{P}\left(\widehat{T}>b\right)=\Phi_1,
\end{align}
where $\Phi_1$ is the rate function defined in
Equation~(\ref{Ia_def}).
\end{lemma}
\begin{remark} We provide a sketch of the proof of
Lemma \ref{lemma_Ia} in the following. Based on
Equation~(\ref{T_power_def}), given any initial state $0\le q<1$,
the event $\widehat{T}(q)>b,b\in\mathbb{N}$ is equivalent to the
event $\sum_{\tau=1}^{t} (\widehat{c}_1[\tau]-\widehat{a}[\tau])\le
q-1,\forall 1\leq t\leq b$. Since
$\widehat{c}_1[\tau]-\widehat{a}[\tau], \forall \tau\in\mathbb{N}$
are \emph{i.i.d.} random variables, the probability of such event
happening at large $b$ can be characterized using large deviation
theories \cite{dembo2010large,weiss1995large}. Then, by combining
with the fact $0\le q<1$, we find the decay rate of
$\mathbb{P}\left(\widehat{T}>b\right)$ that is independent of the
initial states.
%\begin{align}\label{T_exceed upper_1_sketch}
%&\mathbb{P}\left(\widehat{T}>b\right)\approx\nonumber\\&
%\mathbb{P}\left(\sum_{\tau=1}^{t}
%(\widehat{c}_1[\tau]-\widehat{a}[\tau])\le 0,\forall 1\leq t\leq
%b\right).
%\end{align}
%
%Notice that $\widehat{c}_1[\tau]-\widehat{a}[\tau], \forall
%\tau\in\mathbb{N}$ are \emph{i.i.d.} random variables and
%$\mathbb{E}\left[\widehat{c}_1[\tau]-\widehat{a}[\tau]\right]=\gamma_1-\lambda>0$.
%According to the Cramer's Theorem (see Theorem 2.1.24 in
%\cite{dembo2010large}),
%\begin{align}\label{cramer_sketch}
%\mathbb{P}\left(\sum_{\tau=1}^{t}
%(\widehat{c}_1[\tau]-\widehat{a}[\tau])\le
%0\right)=e^{-t\Phi_1+o(t)},
%\end{align}
%where $\Phi_1$ is the rate function defined in
%Equation~(\ref{Ia_def}). According to the Ballot's Theorem (see
%Theorem 3.3 in \cite{weiss1995large}),
%\begin{align}
%&\mathbb{P}\left(\sum_{\tau=1}^{t}
%\left(\widehat{c}_1[\tau]-\widehat{a}[\tau]\right)\le 0,\forall
%1\leq t\leq b\right)=\nonumber\\&e^{-b\Phi_1+o(b)},\nonumber
%\end{align}
%if and only if Equation~(\ref{cramer_sketch}) holds. Hence, the
%decay rate of $\mathbb{P}\left(\widehat{T}>b\right)$ as $b$ goes to
%infinity is obtained.
\end{remark}
\ifreport \iffinal The proof of Lemma \ref{lemma_Ia} is relegated to
\cite{tech_report} due to space limitations. \else
\begin{proof}
See Appendix~\ref{pf_lemma_2}.
\end{proof} \fi
\else The proof of Lemma \ref{lemma_Ia} is relegated to
\cite{tech_report} due to space limitations. \fi

Let us pick $\epsilon\in(0,\Phi_1)$. By the definition of decay
rate, we can find $N_{\epsilon}\in\mathbb{N}$, such that $\forall
b\in \mathbb{N}, b\ge N_{\epsilon}$, we have
\begin{align}\label{PTK_bounds}
e^{-b(\Phi_1+\epsilon)}<\mathbb{P}\left(\widehat{T}>b\right)<e^{-b(\Phi_1-\epsilon)}.
\end{align}

Combining Equations~(\ref{exceed_P_bounds}) and (\ref{PTK_bounds})
yields, for $k$ large enough,
\ifreport
\begin{align}\label{exceed_P_bounds2}
&\mathbb{P}(D_1>k)\le \frac{e^{-k(\Phi_1-\epsilon)}}{\lambda
\mathbb{E}{\big[}\widehat{T}{\big]}}\left(k+\frac{1}{1-e^{-(\Phi_1-\epsilon)}}\right),\nonumber\\
&\mathbb{P}(D_1>k)\ge \frac{e^{-k(\Phi_1+\epsilon)}}{\lambda
\mathbb{E}{\big[}\widehat{T}{\big]}}.
\end{align}
\else
\begin{align}\label{exceed_P_bounds2}
&\frac{e^{-k(\Phi_1+\epsilon)}}{\lambda
\mathbb{E}{\big[}\widehat{T}{\big]}}\le \mathbb{P}(D_1>k)\le
\frac{e^{-k(\Phi_1-\epsilon)}}{\lambda
\mathbb{E}{\big[}\widehat{T}{\big]}}\left(k+\frac{1}{1-e^{-(\Phi_1-\epsilon)}}\right),\mathbb{P}(D_1>k).
\end{align}
\fi On account of $\lim_{k\to\infty}\frac{\log k}{k}=0$,
Equation~(\ref{exceed_P_bounds2}) leads to
\begin{align}
\Phi_1-\epsilon\le-\lim_{k\to\infty}\frac{1}{k}\log
\mathbb{P}(D_1>k)\le \Phi_1+\epsilon.\nonumber
\end{align}
Since $\epsilon$ can be arbitrarily close to 0, the decay rate of decoding delay is
proved.

Note that $f(x)=e^{-\theta x}$ is a convex function. By Jensen's
Inequality, we have $\mathbb{E}\left[e^{-\theta a[t]}\right]\ge
e^{-\theta \mathbb{E}[a[t]]}=e^{-\theta \lambda}$, where the
equality holds when $a[t]=\lambda$. Combining with
Equation~(\ref{Ia_def}), we have
\ifreport
\begin{align}
\Phi_1&\le\sup_{\theta\in\mathbb{R}}\left\{\theta\lambda-\log\left(\gamma_1
e^\theta+1-\gamma_1\right)\right\}\nonumber\\&\stackrel{(a)}{=}-\lambda\log\frac{\lambda}{\gamma_1}+(1-\lambda)\log\frac{1-\lambda}{1-\gamma_1},
\end{align}
\else
\begin{align}
\Phi_1&\le\sup_{\theta\in\mathbb{R}}\left\{\theta\lambda-\log\left(\gamma_1
e^\theta+1-\gamma_1\right)\right\}\stackrel{(a)}{=}-\lambda\log\frac{\lambda}{\gamma_1}+(1-\lambda)\log\frac{1-\lambda}{1-\gamma_1},
\end{align}
\fi where in step (a), the supreme of
$h(\theta)\triangleq\theta\lambda-\log\left(\gamma_1
e^\theta+1-\gamma_1\right)$ can be easily obtained noting that
$h(\theta)$ is a concave function and there is a unique solution of
the equation $\frac{d}{d\theta}h(\theta)=0$.

\section{Proof of Theorem 2}\label{sec:pftheorem2}
In this subsection, we focus on the injection process
$a[t]=\lambda,\forall t$ which would incur the maximum decay rate of
decoding delay. Without loss of generality, we study the average
decoding delay of receiver $1$.

\begin{lemma}\label{lemma_D}
The average decoding delay of receiver $1$ is upper bounded by
\begin{align}\label{avD_deri_2}
&\overline{D}_1\le\frac{1}{2}\frac{\mathbb{E}{\big[}\widehat{T}^2{\big]}}{\mathbb{E}{\big[}\widehat{T}{\big]}}+\frac{5}{2\lambda}.
%&\overline{D}_1\ge\frac{1}{2}\frac{\mathbb{E}\left[T^2\right]}{\mathbb{E}\left[T\right]}-\frac{2\lambda+1}{2\lambda}.
\end{align}
\end{lemma}
\begin{figure}[t]
\centering
\includegraphics[width=3.5in]{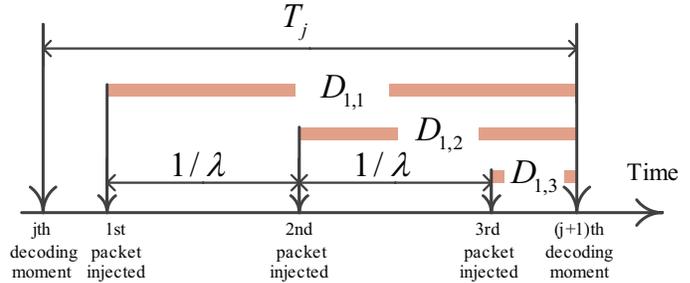}
  \caption{Intuition behind the proof of Lemma \ref{lemma_D}.}
  \label{fig:avD_intuition}
\end{figure}
\begin{remark} The decoding process forms a markov
renewal process \cite{ccinlar1975exceptional}, and the decoding
delay of each packet can be viewed as the residual time from the
epoch when the packet is arrived, till the point when a decoding
happens, which is illustrated in Figure \ref{fig:avD_intuition}.
Then, we can use standard theorem for the markov renewal process to
characterize the average packet decoding delay.
%The proof of Lemma \ref{lemma_D} is intuitive, as shown in Figure
%\ref{fig:avD_intuition}. For the $j^{th}$ decoding interval $T_j$,
%there are approximately $\lambda T_j$ packets decoded. Intutively,
%the decoding delay of these packets form a arithmetic sequence with
%a difference $1/\lambda$. The summation of the delay of all packets
%divided by the number of decoded packets leads to the average delay,
%which is related to both ${T_j}^2$ and $T_j$.
\end{remark}
\ifreport \iffinal The proof of Lemma \ref{lemma_D} is relegated to
\cite{tech_report} due to space limitations. \else
\begin{proof}
See Appendix~\ref{pf_lemma_3}.
\end{proof} \fi
\else The proof of Lemma \ref{lemma_D} is relegated to
\cite{tech_report} due to space limitations. \fi

Hence, it suffices to derive
$\frac{\mathbb{E}{\big[}\widehat{T}^2{\big]}}{\mathbb{E}{\big[}\widehat{T}{\big]}}$.

\begin{lemma}\label{lemma_Y}
Let $Y(q)\triangleq q+\sum_{\tau=1}^{\widehat{T}(q)}
(\widehat{a}[\tau]-\widehat{c}_1[\tau])$. Then, the first two
moments of $\widehat{T}$ can be given by
\ifreport
\begin{align}\label{ETET2}
\mathbb{E}{\big[}\widehat{T}{\big]}&=\frac{\mathbb{E}\left[\mathbb{E}\left[Y(\widehat{Q}_1)|\widehat{Q}_1\right]\right]-\mathbb{E}[\widehat{Q}_1]}{\mu},\nonumber\\
\mathbb{E}{\big[}\widehat{T}^2{\big]}&\le\left(\mathbb{E}\left[\mathbb{E}\left[Y(\widehat{Q}_1)|\widehat{Q}_1\right]\right]-\mathbb{E}[\widehat{Q}_1]\right)\left(\frac{\sigma^2}{\mu^3}-\frac{2}{\mu^2}\right),
%\mathbb{E}\left[T^2\right]&\ge\left(\mathbb{E}[Y]-\mathbb{E}[X]\right)\left(\frac{\sigma^2}{\mu^3}+\frac{2}{\mu^2}\right)-\frac{(2-\lambda)^2}{\mu^2},
\end{align}
\else
\begin{align}\label{ETET2}
\mathbb{E}{\big[}\widehat{T}{\big]}&=\frac{\mathbb{E}\left[\mathbb{E}\left[Y(\widehat{Q}_1)|\widehat{Q}_1\right]\right]-\mathbb{E}[\widehat{Q}_1]}{\mu},\quad\mspace{-13mu}
\mathbb{E}{\big[}\widehat{T}^2{\big]}\le\left(\mathbb{E}\left[\mathbb{E}\left[Y(\widehat{Q}_1)|\widehat{Q}_1\right]\right]-\mathbb{E}[\widehat{Q}_1]\right)\left(\frac{\sigma^2}{\mu^3}-\frac{2}{\mu^2}\right),
%\mathbb{E}\left[T^2\right]&\ge\left(\mathbb{E}[Y]-\mathbb{E}[X]\right)\left(\frac{\sigma^2}{\mu^3}+\frac{2}{\mu^2}\right)-\frac{(2-\lambda)^2}{\mu^2},
\end{align}
\fi where $\mu=\lambda-\gamma_1,\sigma^2=\gamma_1(1-\gamma_1)$ are
the mean and variance of $a[t]-c_1[t]$, respectively.
\end{lemma}
\begin{remark} First, we show that, for any initial
state $q$, $\widehat{T}$ is a stopping time. Using Wald's identity,
we are able to derive the first and second moments of $\widehat{T}$
given the initial state $q$. Then, by combining the fact that $0\le
q<1$, we find the upper bounds for both the first and the second
moments that are independent of the initial states.
%In Lemma \ref{lemma_Y}, $\widehat{T}$ is found to be a stopping time
%if the initial state $q$ is given. Thus, Wald's identities (see
%Theorem 3 in page 488 in \cite{shiryaev1996probability}) can be
%applied to derive the first and second moments of $\widehat{T}$
%given the initial state $q$. Utilizing the fact $0\le q<1$, an upper
%bound of $\mathbb{E}{\big[}\widehat{T}^2{\big]}$ can be derived,
%which is independent of the initial state. Since the distribution of
%$q$ is unknown, Equation~\eqref{ETET2} has a complex form.
\end{remark}
\ifreport \iffinal The proof of Lemma \ref{lemma_Y} is relegated to
\cite{tech_report} due to space limitations. \else
\begin{proof}
See Appendix~\ref{pf_lemma_4}.
\end{proof} \fi
\else The proof of Lemma \ref{lemma_Y} is relegated to
\cite{tech_report} due to space limitations. \fi

\begin{comment}
\begin{lemma}\label{lemma_4}
Let $X\triangleq \widehat{Q}_1$ and $Y\triangleq
\widehat{Q}_1+\sum_{\tau=1}^{\widehat{T}}
(\widehat{a}[\tau]-\widehat{c}_1[\tau])$ denote the state of the
random walk in the beginning and at the end of the $\widehat{T}$
time slots. Then, the first two moments of $\widehat{T}$ can be
given by
\begin{align}\label{ETET2}
\mathbb{E}{\big[}\widehat{T}{\big]}&=\frac{\mathbb{E}[Y]-\mathbb{E}[X]}{\mu},\nonumber\\
\mathbb{E}{\big[}\widehat{T}^2{\big]}&\le\left(\mathbb{E}[Y]-\mathbb{E}[X]\right)\left(\frac{\sigma^2}{\mu^3}-\frac{2}{\mu^2}\right),
%\mathbb{E}\left[T^2\right]&\ge\left(\mathbb{E}[Y]-\mathbb{E}[X]\right)\left(\frac{\sigma^2}{\mu^3}+\frac{2}{\mu^2}\right)-\frac{(2-\lambda)^2}{\mu^2},
\end{align}
where $\mu=\lambda-\gamma_1,\sigma^2=\gamma_1(1-\gamma_1)$ are the
mean, variance of $\chi[t]=a[t]-c_1[t]$ respectively.
\end{lemma}
\end{comment}

From Equation~(\ref{ETET2}),
\ifreport
\begin{align}\label{EN2EN_expand}
\frac{\mathbb{E}{\big[}\widehat{T}^2{\big]}}{\mathbb{E}{\big[}\widehat{T}{\big]}}&\le
\frac{\sigma^2}{\mu^2}-\frac{2}{\mu}
=\frac{\gamma_1(1-\gamma_1)}{(\gamma_1-\lambda)^2}+\frac{2}{\gamma_1-\lambda}.
\end{align}
\else
\begin{align}\label{EN2EN_expand}
\frac{\mathbb{E}{\big[}\widehat{T}^2{\big]}}{\mathbb{E}{\big[}\widehat{T}{\big]}}&\le
\frac{\sigma^2}{\mu^2}-\frac{2}{\mu}
=\frac{\gamma_1(1-\gamma_1)}{(\gamma_1-\lambda)^2}+\frac{2}{\gamma_1-\lambda}.
\end{align}
\fi

Together with Equation~(\ref{avD_deri_2}),
Equation~(\ref{avD_det_bound}) is obtained. It is then
straightforward to see Equation~(\ref{avD_det_limit}).

\section{Proof of Theorem 3}\label{sec:pftheorem3}
According to Equation~(\ref{Wt_bounds}), to get the scaling law of
$W[t]$, it suffices to find the scaling law of $\max_{1\leq i\leq
n}Q_i[t]$. Let $Q_i$ be a random variable with a distribution as the
steady state distribution of $Q_i[t]$. More precisely,
$\mathbb{P}({Q}_i>q)=\mathbb{P}(Q_i[\infty]>q)$ for any $q$. From
Equations~\eqref{decoder_queue} and \eqref{S_def}, we can obtain an
upper bound of $\mathbb{P}({Q}_i>q)$ for any $q$ by letting
$\gamma_i=\gamma$. Together with the fact that $\{Q_i[t]\}_{1\leq
i\leq n}$ are independent, as suggested by Statement
\ref{independency}, it suffices to prove the case when
$\gamma_1=\dots=\gamma_n=\gamma$.

\begin{lemma}\label{lemma_GIGI1}
For an arbitrary receiver $i$ with $\gamma_i=\gamma$,
\begin{align}
-\lim_{k\to\infty} \frac{1}{k}\mathbb{P}(Q_i>k)=\eta,\nonumber
\end{align}
where $\eta$ defined in Equation~(\ref{eq:rate_queue}).
\end{lemma}
\begin{remark} Consider the number of ``unseen''
packets $\lfloor Q_i[t]\rfloor$ for receiver $i$. The number of data
packets that have entered the encoder buffer up to time-slot $t$ is
$\lfloor A[t]\rfloor=\lfloor \lambda t \rfloor$. When $\lfloor
Q_i[t]\rfloor \geq 1$, receiver $i$ has at least one ``unseen''
packet. In this case, the service time for receiver $i$ to see one
more packet is \emph{i.i.d.} geometrically distributed with mean
$1/\gamma$. When $\lfloor Q_i[t]\rfloor = 0$, receiver $i$ needs to
wait for another data packet to enter the encoder buffer before
serving it. We show, through a sample-path argument, that the
evolution of $Q_i[t]$ can be closely characterized by a D/Ge/1 queue
up to a constant difference in the queue length. Then, we can
utilize Proposition 9 in \cite{glynn1993logarithmic} to derive the
delay rate of $Q_i[t]$.
\end{remark}
\ifreport \iffinal The proof of Lemma \ref{lemma_GIGI1} is relegated
to \cite{tech_report} due to space limitations. \else
\begin{proof}
See Appendix~\ref{pf_lemma_5}.
\end{proof} \fi
\else The proof of Lemma \ref{lemma_GIGI1} is relegated to
\cite{tech_report} due to space limitations. \fi

As we discussed in the beginning of Section \ref{sec:proof}, for the
constant injections ($a[t]=\lambda,\forall t$), $\{Q_i[t]\}_{1\leq
i\leq n}$ are independent. Combining with
Equation~(\ref{Wt_bounds}), we need to evaluate the expectation of
the maximum of $n$ \emph{i.i.d.} random variables.

Let us pick $\epsilon\in(0,\eta)$. Then, by Lemma \ref{lemma_GIGI1},
we can find $N_{0}$ such that for any $k\in\mathbb{R},k\ge N_{0}$,
\[e^{-(\eta+\epsilon)k}<\mathbb{P}(Q_i>k)<e^{-(\eta-\epsilon)k},\forall i\in \{1,...,n\}.\]
Introduce two auxiliary random variables $A_U$ and $A_L$ with the following distributions, respectively.
\begin{eqnarray*}
\mathbb{P}(A_U>k)=\left\{
\begin{array}{ll}
1, & \mbox{\normalfont when}\: k\le N_{0}; \\
e^{-(\eta-\epsilon)(k-N_{0})}, &\mbox{\normalfont otherwise},
\end{array}
\right.
\end{eqnarray*}
\begin{eqnarray*}
\mathbb{P}(A_L>k)=\left\{
\begin{array}{ll}
1, & \mbox{\normalfont when}\: k\le N_1; \\
e^{-(\eta+\epsilon)(k-N_1)}, &\mbox{\normalfont otherwise}.
\end{array}
\right.
\end{eqnarray*}
where $N_1=\frac{1}{\eta+\epsilon}\log \mathbb{P}(Q_i>N_0)$.

From $\mathbb{P}(A_L>0)=\mathbb{P}(Q_i>N_0)$,
$\mathbb{P}(A_U>N_0)=1$ and the monotonicity,
\begin{align}
\mathbb{P}(Q_i>N_0)\le\mathbb{P}(Q_i>k)\le 1, \forall k\in [0,N_0],
\nonumber
\end{align}
it can be verified that
\begin{align}\label{P_Qi_bound}
\mathbb{P}(A_L>k)\le \mathbb{P}(Q_i>k)\le \mathbb{P}(A_U>k),\forall
k\in \mathbb{R}.
\end{align}

Let $A_L^i, A_U^i,i=1,...,n$ be independent random variables with
same distribution as $A_L, A_U$, respectively. Then from
Equation~(\ref{P_Qi_bound}), we have
\begin{align}\label{EmaxQ_bound}
\mathbb{E}\left[\max_{1\leq i\leq n} A_L^i\right]\le
\mathbb{E}\left[\max_{1\leq i\leq n}Q_i\right]\le
\mathbb{E}\left[\max_{1\leq i\leq n} A_U^i\right].
\end{align}

\ifreport The upper and lower bounds in the above equation
correspond to the maximum of $n$ \emph{i.i.d.} exponential random
variables, the expectation of which can be easily calculated
\cite{Emax_exponential}. $\mathbb{E}\left[\max_{1\leq i\leq n}
A_L^i\right]=N_1+\frac{H_n}{\eta+\epsilon}$ and
$\mathbb{E}\left[\max_{1\leq i\leq n}
A_U^i\right]=N_0+\frac{H_n}{\eta-\epsilon}$, in which
$H_n=\sum_{j=1}^n 1/j$ is the harmonic number. \else The upper and
lower bounds in the above equation correspond to the maximum of $n$
\emph{i.i.d.} exponential random variables, the expectation of which
can be easily calculated \cite{Emax_exponential}.
$\mathbb{E}\left[\max_{1\leq i\leq n} A_L^i\right]$
$=N_1+\frac{H_n}{\eta+\epsilon}$ and $\mathbb{E}\left[\max_{1\leq
i\leq n} A_U^i\right]=N_0+\frac{H_n}{\eta-\epsilon}$, in which
$H_n=\sum_{j=1}^n 1/j$ is the harmonic number. \fi By taking the
expectation of Equation~(\ref{Wt_bounds}), we have
\begin{align}\label{EW_bounds}
\mathbb{E}\left[\max_{1\leq i\leq n} Q_i\right]-1\le
\overline{W}_n\le \mathbb{E}\left[\max_{1\leq i\leq n} Q_i\right]+1,
\end{align}
which, together with the fact that $\lim_{n\to\infty} H_n/{\log n}=1$, yields
\[\frac{1}{\eta+\epsilon}\le\lim_{n\to\infty}\frac{\overline{W}_n}{\log
n}\le\frac{1}{\eta-\epsilon}.\] Since $\epsilon$ can be arbitrarily
close to 0, Equation~(\ref{avW_order}) is derived.

Next, we prove the decay rate of encoding complexity for a fixed number of receivers $n$.

From Equation~(\ref{P_Qi_bound}), we have, for any $k\in
\mathbb{R}$,\ifreport
\begin{align}\label{P_maxQ_bound}
&\mathbb{P}\left(\max_{1\leq i\leq n}Q_i>k\right)\le
\mathbb{P}\left(\max_{1\leq i\leq n}A_U^i>k\right),\\
&\mathbb{P}\left(\max_{1\leq i\leq n}Q_i>k\right)\ge\mathbb{P}\left(\max_{1\leq i\leq n}A_L^i>k\right).\label{P_maxQ_bound1}
\end{align}
\else
\begin{align}\label{P_maxQ_bound}
&\mathbb{P}\left(\max_{1\leq i\leq
n}A_L^i>k\right)\le\mathbb{P}\left(\max_{1\leq i\leq
n}Q_i>k\right)\le \mathbb{P}\left(\max_{1\leq i\leq
n}A_U^i>k\right).
\end{align}
\fi

According to Proposition~3.2 in \cite{bi2013}, the complementary
cumulative distribution function of the maximum of independent
exponentially distributed variables $\left\{A_U^i\right\}_{1\le i\le
n}$ is given by\ifreport
\begin{align}
&\mathbb{P}\left(\max_{1\leq i\leq n}A_U^i>k\right)=\sum_{i=1}^n
(-1)^{i+1}\binom{n}{i}e^{-i(\eta-\epsilon)(k-N_0)}\nonumber\\
&=e^{-(\eta-\epsilon)(k-N_0)}\left(n+\sum_{i=2}^n
(-1)^{i+1}\binom{n}{i}e^{-(i-1)(\eta-\epsilon)(k-N_0)}\right)\nonumber\\
&=e^{-(\eta-\epsilon)(k-N_0)}(n+o(1)),\nonumber
\end{align}
\else
\begin{align}
&\mathbb{P}\left(\max_{1\leq i\leq n}A_U^i>k\right)=\sum_{i=1}^n
(-1)^{i+1}\binom{n}{i}e^{-i(\eta-\epsilon)(k-N_0)}\nonumber\\
=& e^{-(\eta-\epsilon)(k-N_0)}\left(n+\sum_{i=2}^n
(-1)^{i+1}\binom{n}{i}e^{-(i-1)(\eta-\epsilon)(k-N_0)}\right)=
e^{-(\eta-\epsilon)(k-N_0)}(n+o(1)),\nonumber
\end{align}
\fi where $o(1)$ converges to $0$ as $k\to\infty$. By combining the
above equation with Equation~(\ref{P_maxQ_bound}), we have,\ifreport
\begin{align}
&-\lim_{k\to\infty}\frac{1}{k}\log\mathbb{P}\left(\max_{1\leq i\leq
n}Q_i>k\right)\ge\nonumber\\&-\lim_{k\to\infty}\frac{1}{k}\log\left(e^{-(\eta-\epsilon)(k-N_0)}(n+o(1))\right)=\eta-\epsilon.\notag
\end{align}
\else
\begin{align}
&-\lim_{k\to\infty}\frac{1}{k}\log\mathbb{P}\left(\max_{1\leq i\leq
n}Q_i>k\right)\ge-\lim_{k\to\infty}\frac{1}{k}\log\left(e^{-(\eta-\epsilon)(k-N_0)}(n+o(1))\right)=\eta-\epsilon.\notag
\end{align}
\fi

\ifreport The other direction can be proven using the same procedure
on Equation~(\ref{P_maxQ_bound1}). It is clear from
Equation~(\ref{Wt_bounds}) that $\mathbb{P}\left(W_n>k\right)$ has
the same decay rate as $\mathbb{P}\left(\max_{1\leq i\leq
n}Q_i>k\right)$, thus Equation~(\ref{W_decay_rate}) is proved. \else
The other direction can be proven using the same procedure starting
from the other half of Equation~(\ref{P_maxQ_bound}). It is clear
from Equation~(\ref{Wt_bounds}) that $\mathbb{P}\left(W_n>k\right)$
has the same decay rate as $\mathbb{P}\left(\max_{1\leq i\leq
n}Q_i>k\right)$, thus Equation~(\ref{W_decay_rate}) is proved. \fi

\section{Proof of Theorem 4}\label{sec:pftheorem4}
Similar to the proof of Theorem \ref{theorem3}, we prove for the
case when $\gamma_1=\dots=\gamma_n=\gamma$.

Without loss of generality, we focus on receiver $1$. Take one time
of addition and multiplication as one operation. Let $Q_i$ be a
random variable whose distribution is the same as the steady state
distribution of $Q_i[t]$.

%In the first place, let's check the decoding of a number of $K_j$
%packets in the $j^\text{th}$ decoding interval $T_j$. Note that the
%encoding window size $W[t]$ depends on all the receivers in the
%network. To upper bound the decoding complexity, we need to

\begin{figure*}[t]
\centering
\includegraphics[width=0.95\textwidth]{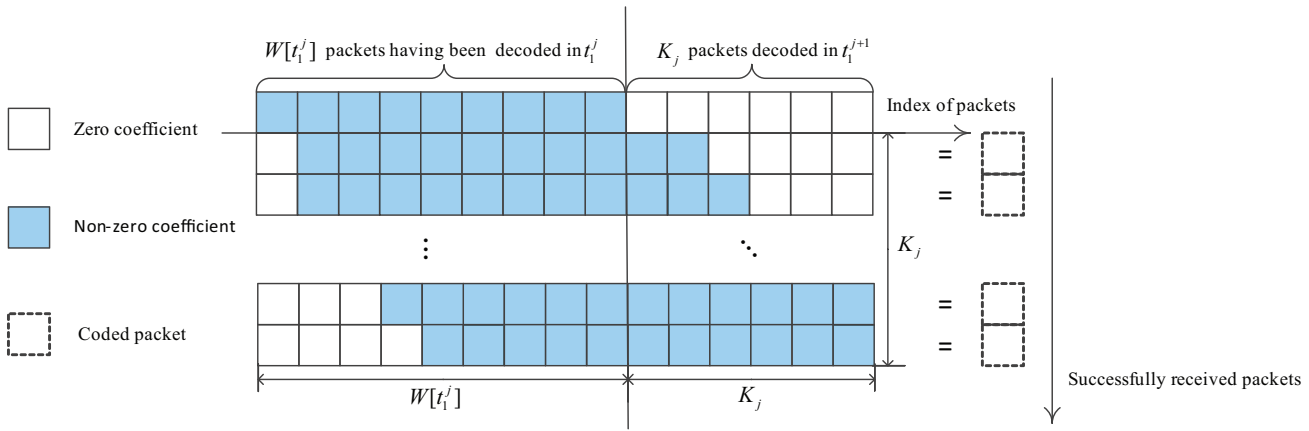}
  \caption{Intuition behind the proof of Theorem \ref{theorem4}.}
  \label{fig:Decode_intuition}
\end{figure*}

At time-slot $t_1^j$, all the packets in the encoder buffer
$W[t_1^j]$ have been decoded at receiver $i$. Then, after a decoding
interval of $T_j$, at time-slot $t_1^{j+1}$, $K_j$ more packets are
decoded at receiver $i$, as shown in Figure
\ref{fig:Decode_intuition}. To upper bound the decoding complexity,
we need an upper bound of $W[t]$ for each time-slot
$t\in(t_1^j,t_1^j+T_j]$. An obvious upper bound is $W[t_1^j]+K_j\geq
W[t]$. Thus, each coded packet received within the interval
$(t_1^j,t_1^j+T_j]$ can be encoded from at most a number of
$W[t_1^j]+K_j$ packets. As a result, the coefficients of the $K_j$
received coded packets can form a decoding matrix with $K_j$ rows
and $W[t_1^j]+K_j$ columns, where each row corresponds to a data
packet and each column corresponds to a coded packet received. We
categorize the decoding process into two steps.

\begin{itemize}[\setlabelwidth{0em}]
\item[Step 1:] Since the packets corresponding to the first
$W[t_1^j]$ columns have been decoded in slot $t_1^j$, the receiver
could apply a maximum number of $K_j W[t_1^j]$ operations so that
the $K_j\times\left(W[t_1^j]+ K_j\right)$ matrix is reduced to a $
K_j\times K_j$ matrix.
\item[Step 2:] Gauss-Jordan elimination is performed to decode
from the reduced matrix which takes $O\left(( K_j)^3\right)$
operations.
\end{itemize}

In the following we derive the average decoding complexity taken by
Step 1 and Step 2 respectively.

\begin{lemma}\label{lemma_OmegaW}
Let $\overline{\Omega}_{n,1}$ denote the average complexity taken by
Step 1 to decode a packet, then

\begin{align}
\overline{\Omega}_{n,1}\le\overline{W}_n+\overline{C}_U,
\end{align}
in which $\overline{W}_n$ denotes the average encoding complexity,
and $\overline{C}_U$ is a constant independent of $n$.
\end{lemma}
\begin{remark} The motivation of Lemma
\ref{lemma_OmegaW} is the following. In Step 1, at most $K_j
W[t_1^j]$ operations are needed for the $K_j$ packets to be decoded.
As a result, the average decoding complexity for each packet in Step
1 is upper bounded by $W[t_1^j]$, which scales in the same order as
$\overline{W}_n$ as $n$ increases.
\end{remark}
\ifreport \iffinal The proof of Lemma \ref{lemma_OmegaW} is
relegated to \cite{tech_report} due to space limitations. \else
\begin{proof}
See Appendix~\ref{pf_lemma_6}.
\end{proof} \fi
\else The proof of Lemma \ref{lemma_OmegaW} is relegated to
\cite{tech_report} due to space limitations. \fi

For ease of presentation, we assume there exists a constant
$M_\text{C}$ such that Gauss elimination for $m$ packets in Step 2
takes at most $M_\text{C}m^3$ operations.
\begin{lemma}\label{lemma_Omega}
Let $\overline{\Omega}_{n,2}$ denote the average complexity taken by
Step 2 to decode a packet, then
\begin{align}
\overline{\Omega}_{n,2}\le\frac{M_\text{C}\mathbb{E}\left[\mathbb{E}\left[\left(\lambda
\widehat{T}(\widehat{Q}_1)+1\right)^3\big|\widehat{Q}_1\right]\right]}{\lambda
\mathbb{E}{\big[}\widehat{T}{\big]}}.
\end{align}
\end{lemma}
\begin{remark} For the $K_j$ packets to be decoded,
Step 2 takes at most $M_\text{C}{K_j}^3$ operations. For constant
data injection process, given the initial state $Q_1[t_1^j]$, $K_j$
and $T_j$ uniquely determine each other. Thus, it is possible to
upper bound the decoding complexity taken by Step 2 by expressions
only involving the decoding intervals $\{T_j\}_{j=1,\dots}$. Since
the decoding process forms a markov renewal process, applying
standard theorem for the markov renewal process leads to Lemma
\ref{lemma_Omega}.
\end{remark}
\ifreport \iffinal The proof of Lemma \ref{lemma_Omega} is relegated
to \cite{tech_report} due to space limitations. \else
\begin{proof}
See Appendix~\ref{pf_lemma_7}.
\end{proof} \fi
\else The proof of Lemma \ref{lemma_Omega} is relegated to
\cite{tech_report} due to space limitations. \fi

The aggregate decoding complexity is the sum of the complexity by
Step 1 and Step 2. Thus, \ifreport
\begin{align}\label{Omega_upper}
\overline{\Omega}_n&=\overline{\Omega}_{n,1}+\overline{\Omega}_{n,2}\nonumber\\
&\le
\overline{W}_n+\overline{C}_U+\frac{M_\text{C}\mathbb{E}\left[\mathbb{E}\left[\left(\lambda
\widehat{T}(\widehat{Q}_1)+1\right)^3\big|\widehat{Q}_1\right]\right]}{\lambda
\mathbb{E}{\big[}\widehat{T}{\big]}}.
\end{align}
\else
\begin{align}\label{Omega_upper}
\overline{\Omega}_n&=\overline{\Omega}_{n,1}+\overline{\Omega}_{n,2}\le
\overline{W}_n+\overline{C}_U+\frac{M_\text{C}\mathbb{E}\left[\mathbb{E}\left[\left(\lambda
\widehat{T}(\widehat{Q}_1)+1\right)^3\big|\widehat{Q}_1\right]\right]}{\lambda
\mathbb{E}{\big[}\widehat{T}{\big]}}.
\end{align}
\fi

By Lemma~\ref{lemma_Ia}, $\mathbb{P}\left(\widehat{T}>k\right)$
decays exponentially for large enough $k$, thus
$\mathbb{E}\left[\widehat{T}^3\right]$ is finite. It can be ``seen''
in Equation~\eqref{T_hat} that for given $\lambda$, the distribution
of $\widehat{T}$ is independent of the number of receivers $n$, thus
the last term in Equation~\eqref{Omega_upper} remains unchanged for
arbitrarily large $n$.

To find the lower bound of the average decoding complexity, we have
the following lemma.
\begin{lemma}\label{lemma_Omega_lower}
The average decoding complexity of MWNC-AF is lower bounded by the
average encoding complexity of MWNC-AF.
\begin{align}\label{Omega_lower}
\overline{\Omega}_n\ge\overline{W}_{n-1}-\overline{C}_L,
\end{align}
in which $\overline{W}_{n-1}$ denotes the average encoding
complexity given there are $n-1$ receivers, and $\overline{C}_L$ is
a constant independent of $n$.
\end{lemma}
\ifreport \iffinal The proof of Lemma \ref{lemma_Omega_lower} is
relegated to \cite{tech_report} due to space limitations. \else
\begin{proof}
See Appendix~\ref{pf_lemma_8}.
\end{proof} \fi
\else The proof of Lemma \ref{lemma_Omega_lower} is relegated to
\cite{tech_report} due to space limitations. \fi

From Equation~\eqref{Omega_upper} and \eqref{Omega_lower}, we could
have
\begin{align}
\lim_{n\to\infty}\frac{\overline{\Omega}_n}{\log
n}=\lim_{n\to\infty}\frac{\overline{W}_{n-1}}{\log
n}=\lim_{n\to\infty}\frac{\overline{W}_n}{\log n}.\nonumber
\end{align}
With Equation~(\ref{avW_order}), Equation~(\ref{avOmega_order}) is
proved.

\ifreport \iffinal \else
\section{Proof for Lemma~\ref{lem_feedback}}\label{pf_lemma_0}
We prove Lemma \ref{lem_feedback} by induction. In time-slot $0$,
this is true because $Z[0]=Z_i[0]=\min_{1\leq i\leq n}S_i[0]=0$.
Suppose that
\begin{eqnarray}\label{eq_state}
Z[t-1]=Z_i[t-1]=\min_{1\leq i\leq n} S_i[t-1]
\end{eqnarray}
is satisfied at the end of time-slot $t-1$. If there exists some
receiver $i$ that satisfies $S_i[t]=Z_i[t-1]$, then we have
$\min_{1\leq i\leq n} S_i[t]=\min_{1\leq i\leq n} S_i[t-1]$. By
Lines 8 and 25 of Algorithm \ref{feedbackalg}, the transmitter can
detect a beacon signal such that $Z[t] =Z[t-1]$. Otherwise, if
$S_i[t]\neq Z_i[t-1]$ for each receiver $i$, then by Algorithm
\ref{feedbackalg}, the transmitter will detect no beacon signal such
that $Z[t] =Z[t-1] + 1$. Meanwhile,
Equations~\eqref{S_def},~\eqref{eq_state}, and $S_i[t]\neq Z_i[t-1]$
tell us that $\min_{1\leq i\leq n} S_i[t]=\min_{1\leq i\leq n}
S_i[t-1]+1$. Since $Z_i[t]$ with $Z[t]$ are synchronized, we have
\begin{eqnarray}
Z[t]=Z_i[t]=\min_{1\leq i\leq n} S_i[t]\nonumber
\end{eqnarray}
for time-slot $t$.

\section{Proof for Lemma~\ref{lemma_D1k}}\label{pf_lemma_1}

Let ${K}_j^k$ denote the number of packets with decoding delay
greater than the threshold $k$ for the decoding interval ${T}_j,j\in
\mathbb{N}$. Analogous to Equation~\eqref{K_j_def}, ${K}_j^k$ can be
given by
\begin{align}\label{Kjk_def}
K_j^k=\left\lfloor Q_1[t_1^j]+
\sum_{t=t_1^{j}+1}^{t_1^{j}+T_j-k}a[t]\right\rfloor.
\end{align}

By the definition of delay exceeding probability (given by
Equation~(\ref{exceed_P_def})), the numerator can be expressed as
the sum of the number of packets exceeding the threshold in the
decoding intervals,
\begin{align}\label{exceed_P_def2}
\mathbb{P}(D_1>k)&=\lim_{J\to\infty}\frac{\sum_{j=1}^J
{K}_j^k}{\sum_{j=1}^J {K}_j}\nonumber\\
&=\lim_{J\to\infty}\frac{\sum_{j=1}^{J} {K}_j^k} {\sum_{j=1}^{J}
{T}_j}\cdot\lim_{J\to\infty}\frac{\sum_{j=1}^{J}
{T}_j}{\sum_{j=1}^{J} {K}_j}.
\end{align}
Subsequently, we show how to derive the properties for the two limit
terms on the right side of Equation~(\ref{exceed_P_def2}).

The second limit term is simple. By
Equation~\eqref{def_decoding_moment}, at a decoding moment $t_1^j$,
all packets up to $\lfloor A[t_1^j]\rfloor$ are decoded by receiver
1. If $t=\sum_{j=1}^J {T}_j$, with Equation~(\ref{At_def}) we have
\begin{align}\label{lim_trivial}
\lim_{J\to\infty}\frac{\sum_{j=1}^{J} {T}_j}{\sum_{j=1}^{J}
{K}_j}=\lim_{t\to\infty}\frac{t}{\lfloor \sum_{\tau=1}^t
a[\tau]\rfloor}\stackrel{(a)}{=}\frac{1}{\lambda},
\end{align}
where in step (a), strong law of large numbers is applied on
\emph{i.i.d.} random variables $a[\tau],\forall \tau$.

To bound the first limit term in Equation~(\ref{exceed_P_def2}), we
observe the following facts for the packets decoded after the
interval $ {T}_j$, which can be seen from
Equations~\eqref{T_power_def} and \eqref{Kjk_def}.
\begin{enumerate}
\item If $ {T}_j\le k$, there is no packets exceeding the threshold $k$, i.e., $ {K}_j^k=0$.
\item If $ {T}_j> k$, there are at most $ {T}_j$ packets which exceed the threshold $k$, i.e., $ {K}_j^k\le {T}_j$.
\item If $ {T}_j> k$, there is at least one packet which exceed the threshold $k$, i.e., $ {K}_j^k\ge 1$.
\end{enumerate}

Thus,
\begin{align}\label{1st_limit_bounds}
\lim_{J\to\infty}\frac{\sum_{j=1}^{J} {K}_j^k} {\sum_{j=1}^{J}
{T}_j}&\le\lim_{J\to\infty}\frac{\sum_{j=1}^{J}1_{\left\{
{T}_j>k\right\}} {T}_j} {\sum_{j=1}^{J}
{T}_j},\nonumber\\\lim_{J\to\infty}\frac{\sum_{j=1}^{J} {K}_j^k}
{\sum_{j=1}^{J}
{T}_j}&\ge\lim_{J\to\infty}\frac{\sum_{j=1}^{J}1_{\left\{
{T}_j>k\right\}}} {\sum_{j=1}^{J} {T}_j}.
\end{align}

Consider $1_{\left\{ {T}_j>k\right\}} {T}_j$ and $1_{\left\{
{T}_j>k\right\}}$ as the rewards earned in interval $ {T}_j$.
According to the Markov renewal reward theory (see Theorem 11.4
\cite{ccinlar1975exceptional}), we have
\begin{align}\label{exceed_P_renewal}
&\lim_{J\to\infty}\frac{\sum_{j=1}^{J}1_{\left\{{T}_j>k\right\}}
{T}_j}
{\sum_{j=1}^{J}{T}_j}=\frac{\mathbb{E}\left[\mathbb{E}\left[1_{\left\{\widehat{T}(\widehat{Q}_1)>k\right\}}\widehat{T}(\widehat{Q}_1)\Big|\widehat{Q}_1\right]\right]}{\mathbb{E}\left[\mathbb{E}\left[\widehat{T}(\widehat{Q}_1)\Big|\widehat{Q}_1\right]\right]},\nonumber\\
&\lim_{J\to\infty}\frac{\sum_{j=1}^{J}1_{\left\{
{T}_j>k\right\}}}{\sum_{j=1}^{J}
{T}_j}=\frac{\mathbb{E}\left[\mathbb{E}\left[1_{\left\{\widehat{T}(\widehat{Q}_1)>k\right\}}\Big|\widehat{Q}_1\right]\right]}{\mathbb{E}\left[\mathbb{E}\left[\widehat{T}(\widehat{Q}_1)\Big|\widehat{Q}_1\right]\right]},
\end{align}
where $\widehat{T}(.)$ and $\widehat{Q}_1$ are defined in
Appendix~\ref{sec:proof}. Note that, for any $0\leq q< 1$, we have
\begin{align}
&\mathbb{E}\left[1_{\left\{\widehat{T}(q)>k\right\}}\widehat{T}(q)\right]=\sum_{b=k+1}^\infty
b\mathbb{P}\left(\widehat{T}(q)=b\right)\nonumber\\&\;\;\;\;\;\;\;\;\;=k\mathbb{P}\left(\widehat{T}(q)>k\right)+\sum_{b=k}^\infty
\mathbb{P}\left(\widehat{T}(q)>b\right),\nonumber\\
&\mathbb{E}\left[1_{\left\{\widehat{T}(q)>k\right\}}\right]=\mathbb{P}\left(\widehat{T}(q)>k\right),
\end{align}
which, by combining with Equations~(\ref{exceed_P_def2}),
(\ref{lim_trivial}), (\ref{1st_limit_bounds}) and
(\ref{exceed_P_renewal}), completes the proof of
Equation~(\ref{exceed_P_bounds}).

\section{Proof for Lemma~\ref{lemma_Ia}}\label{pf_lemma_2}
Based on Equation~(\ref{T_power_def}) and Equation~(\ref{PT_hat}),
$\mathbb{P}{\big(}\widehat{T}>b{\big)},b\in\mathbb{N}$ can be expressed as
\begin{align}\label{PTk_def}
&\mathbb{P}\left(\widehat{T}>b\right)=\nonumber\\
&\mathbb{E}\left[\mathbb{P}
\left(\sum_{\tau=1}^{t}
(\widehat{c}_1[\tau]-\widehat{a}[\tau])\le
\widehat{Q}_1-1,\forall 1\leq t\leq b {\Big|}\widehat{Q}_1\right)\right].
\end{align}
Since $\widehat{Q}_1-1<0$, Equation~(\ref{PTk_def}) can be upper
bounded by

\begin{align}\label{T_exceed upper_1}
&\mathbb{P}\left(\widehat{T}>b\right)\le\nonumber\\&
\mathbb{P}\left(\sum_{\tau=1}^{t}
(\widehat{c}_1[\tau]-\widehat{a}[\tau])\le 0,\forall
1\leq t\leq b\right).
\end{align}

Notice that $\widehat{c}_1[\tau]-\widehat{a}[\tau], \forall
\tau\in\mathbb{N}$ are \emph{i.i.d.} random variables and
$\mathbb{E}\left[\widehat{c}_1[\tau]-\widehat{a}[\tau]\right]=\gamma_1-\lambda>0$.
According to the Cramer's Theorem (see Theorem 2.1.24 in
\cite{dembo2010large}),
\begin{align}\label{cramer}
\mathbb{P}\left(\sum_{\tau=1}^{t}
(\widehat{c}_1[\tau]-\widehat{a}[\tau])\le
0\right)=e^{-t\Phi_1+o(t)},
\end{align}
where $\Phi_1$ is the rate function defined in
Equation~(\ref{Ia_def}). According to the Ballot's Theorem (see
Theorem 3.3 in \cite{weiss1995large}),
\begin{align}
&\mathbb{P}\left(\sum_{\tau=1}^{t}
\left(\widehat{c}_1[\tau]-\widehat{a}[\tau]\right)\le 0,\forall
1\leq t\leq b\right)=\nonumber\\&e^{-b\Phi_1+o(b)},\nonumber
\end{align}
if and only if Equation~(\ref{cramer}) holds. Hence, a lower bound
for the decay rate of $\mathbb{P}\left(\widehat{T}>b\right)$ as $b$
goes to infinity is obtained.
\begin{equation}\label{interval_rate_lower}
-\lim_{b\to\infty}\frac{1}{b}\log
\mathbb{P}\left(\widehat{T}>b\right)\ge \Phi_1.
\end{equation}

To prove the other direction, let us define the event
$\widehat{\mathcal{A}}(q)=\left\{\widehat{c}_1[1]-\widehat{a}[1]\le
q-1\right\}$ for each $0\leq q<1$, then from
Equation~(\ref{PTk_def}), we have, \ifreport
\begin{align}
&\mathbb{P}\left(\widehat{T}>b\right)\nonumber\\
=&\mathbb{E}\left[\mathbb{P} \left(\sum_{\tau=1}^{t}
(\widehat{c}_1[\tau]-\widehat{a}[\tau])\le
\widehat{Q}_1-1,\forall 1\leq t\leq b {\Big|}\widehat{Q}_1\right)\right]\notag\\
\geq&\mathbb{E}\left[\mathbb{P} \left(\sum_{\tau=1}^{t}
(\widehat{c}_1[\tau]-\widehat{a}[\tau])\le
\widehat{Q}_1-1,\forall 2\leq t\leq b {\Big|}\widehat{Q}_1,\widehat{\mathcal{A}}(\widehat{Q}_1)\right)\right.\notag\\
&\cdot\mathbb{P}\left(\widehat{\mathcal{A}}(\widehat{Q}_1)\right)\Bigg]\notag\\
=&\mathbb{E}\left[\mathbb{P} \left(\sum_{\tau=2}^{t}
(\widehat{c}_1[\tau]-\widehat{a}[\tau])\le
\widehat{Q}_1-1-\left(\widehat{c}_1[1]-\widehat{a}[1]\right),\right.\right.\notag\\&\forall
2\leq t\leq b
{\Big|}\widehat{Q}_1,\widehat{\mathcal{A}}(\widehat{Q}_1)\Bigg)
\cdot\mathbb{P}\left(\widehat{\mathcal{A}}(\widehat{Q}_1)\right)\Bigg]\notag\\
\geq&\mathbb{E}\left[\mathbb{P} \left(\sum_{\tau=2}^{t}
(\widehat{c}_1[\tau]-\widehat{a}[\tau])\le
0,\forall 2\leq t\leq b {\Big|}\widehat{Q}_1\right)\mathbb{P}\left(\widehat{\mathcal{A}}(\widehat{Q}_1)\right)\right]\notag\\
=&\mathbb{P} \left(\sum_{\tau=2}^{t}
(\widehat{c}_1[\tau]-\widehat{a}[\tau])\le
0,\forall 2\leq t\leq b\right)\mathbb{E}\left[\mathbb{P}\left(\widehat{\mathcal{A}}(\widehat{Q}_1)\right)\right]\notag\\
\stackrel{(a)}{=}&e^{-(b-1)\Phi_1+o(b)}\times\mathbb{E}\left[\mathbb{P}\left(\widehat{\mathcal{A}}(\widehat{Q}_1)\right)\right],\nonumber
\end{align}
\else
\begin{align}
&\mathbb{P}\left(\widehat{T}>b\right)\nonumber\\
=&\mathbb{E}\left[\mathbb{P} \left(\sum_{\tau=1}^{t}
(\widehat{c}_1[\tau]-\widehat{a}[\tau])\le
\widehat{Q}_1-1,\forall 1\leq t\leq b {\Big|}\widehat{Q}_1\right)\right]\notag\\
\geq&\mathbb{E}\left[\mathbb{P} \left(\sum_{\tau=1}^{t}
(\widehat{c}_1[\tau]-\widehat{a}[\tau])\le
\widehat{Q}_1-1,\forall 1\leq t\leq b {\Big|}\widehat{Q}_1,\widehat{\mathcal{A}}(\widehat{Q}_1)\right)\mathbb{P}\left(\widehat{\mathcal{A}}(\widehat{Q}_1)\right)\right]\notag\\
=&\mathbb{E}\left[\mathbb{P} \left(\sum_{\tau=2}^{t}
(\widehat{c}_1[\tau]-\widehat{a}[\tau])\le
0,\forall 2\leq t\leq b {\Big|}\widehat{Q}_1\right)\mathbb{P}\left(\widehat{\mathcal{A}}(\widehat{Q}_1)\right)\right]\notag\\
=&\mathbb{P} \left(\sum_{\tau=2}^{t}
(\widehat{c}_1[\tau]-\widehat{a}[\tau])\le
0,\forall 2\leq t\leq b\right)\mathbb{E}\left[\mathbb{P}\left(\widehat{\mathcal{A}}(\widehat{Q}_1)\right)\right]\notag\\
\stackrel{(a)}{=}&e^{-(b-1)\Phi_1+o(b)}\times\mathbb{E}\left[\mathbb{P}\left(\widehat{\mathcal{A}}(\widehat{Q}_1)\right)\right],\nonumber
\end{align}
\fi
where in step (a), Ballot Theorem is applied. Since the second
term in the above equation is a constant, it follows that
\begin{align}\label{interval_rate_upper}
-\lim_{b\to\infty}\frac{1}{b}\log
\mathbb{P}\left(\widehat{T}>b\right)\le \Phi_1.
\end{align}

Combining Equations~(\ref{interval_rate_lower}) and
(\ref{interval_rate_upper}), we get the decay rate regarding the
decoding interval $T$, Equation~(\ref{interval_rate_final}) is
proved.

\section{Proof for Lemma~\ref{lemma_D}}\label{pf_lemma_3}

Since $a[t]=\lambda,\forall t$, from Equation~\eqref{K_j_def}, for
any decoding interval $ {T}_j$, we have
\begin{align}\label{K_bounds}
\lambda {T}_j-1\le {K}_j=\left\lfloor Q_1[t_1^j]+
{T}_j\lambda\right\rfloor\le \lambda {T}_j+1.
\end{align}

By the definition of decoding delay in Section
\ref{sec:system_model}, the ${K}_j$ packets decoded after the
decoding interval ${T}_j$ may have different decoding delay
depending on the time slot the packets get injected into the
encoder. Notice that for the constant injection process, the packets
can be considered to arrive one by one with a fixed interval
$1/\lambda$. Thus, it is easy to verify that the decoding delay of
the $m^\text{th}$ admitted packet among the ${K}_j$ decoded packets
is upper bounded by ${T}_j-(m-1)/{\lambda}$. Together with
Equation~(\ref{K_bounds}), the sum of the decoding delay of packets
decoded after the interval ${T}_j$ is bounded by
\begin{align}
\sum_{m=1}^{ {K}_j}D_{1,m}&\le\sum_{m=1}^{K_j}
\left(T_j-\frac{m-1}{\lambda}\right)=K_jT_j-\frac{1}{\lambda}\sum_{l=0}^{K_j-1}{l}\nonumber\\
&\le\left(\lambda T_j+1\right)T_j-\frac{\left(\lambda
T_j-1\right)\left(\lambda
T_j-2\right)}{2\lambda}\nonumber\\
&\le
\frac{\lambda}{2}(T_j)^2+\frac{5}{2}T_j-\frac{1}{\lambda}.\notag
%&\sum_{m=1}^{ {K}_j}D_{1,m}\ge\frac{({T}_j-2)(\lambda {T}_j-1)}{2}.
\end{align}
Combining the upper bound with Equation~(\ref{avD_def}), the average
decoding delay of receiver $1$ can be upper bounded by
\begin{align}\label{avD_deri_1}
\overline{D}_1&\le\lim_{J\to\infty}\frac{\sum_{j=1}^J
\left(\frac{\lambda}{2}(T_j)^2+\frac{5}{2}T_j-\frac{1}{\lambda}\right)}{\sum_{j=1}^J {K}_j}\nonumber\\
&=\lim_{J\to\infty}\frac{\sum_{j=1}^{J}\left(\frac{\lambda}{2}(T_j)^2+\frac{5}{2}T_j-\frac{1}{\lambda}\right)}
{\sum_{j=1}^{J} {T}_j}\cdot\lim_{J\to\infty}\frac{\sum_{j=1}^{J}
{T}_j}{\sum_{j=1}^{J} {K}_j},
\end{align}
where the latter limit has been given by
Equation~(\ref{lim_trivial}).

Take $\frac{\lambda}{2}(T_j)^2+\frac{5}{2}T_j-\frac{1}{\lambda}$ as
the reward earned in interval $ {T}_j$. According to the Markov
renewal reward theory (see Theorem 11.4
\cite{ccinlar1975exceptional}), we have
\begin{align}
&\lim_{J\to\infty}\frac{\sum_{j=1}^{J}\left(\frac{\lambda}{2}(T_j)^2+\frac{5}{2}T_j-\frac{1}{\lambda}\right)}{\sum_{j=1}^{J} {T}_j}\notag\\
=&\frac{\mathbb{E}\left[\mathbb{E}\left[\frac{\lambda}{2}\left(\widehat{T}(\widehat{Q}_1)\right)^2+\frac{5}{2}\widehat{T}(\widehat{Q}_1)-\frac{1}{\lambda}\bigg|\widehat{Q}_1\right]\right]}{\mathbb{E}\left[\mathbb{E}\left[\widehat{T}(\widehat{Q}_1)\Big|\widehat{Q}_1\right]\right]}\nonumber\\
\le&\frac{\frac{\lambda}{2}\mathbb{E}\left[\widehat{T}^2\right]+\frac{5}{2}\mathbb{E}\left[\widehat{T}\right]}{\mathbb{E}\left[\widehat{T}\right]},\notag
\end{align}
where $\widehat{T}$ and $\widehat{Q}_1$ are defined in
Appendix~\ref{sec:proof}. The above equation, together with
Equations~(\ref{lim_trivial}) and Equation~(\ref{avD_deri_1}),
completes the proof of Equation~(\ref{avD_deri_2}).

\section{Proof for Lemma~\ref{lemma_Y}}\label{pf_lemma_4}
To begin with, we derive $\mathbb{E}{\big[}\widehat{T}(q){\big]}$
and $\mathbb{E}{\big[}\widehat{T}^2(q){\big]}$ for any $0\leq q<1$.
From Equation~(\ref{T_hat}), we can see that $\widehat{T}(q)=t$ only
depends on the realizations of
$\left\{\widehat{a}[\tau],\widehat{c}[\tau]\right\}_{1\le \tau\le
t}$, thus $\widehat{T}(q)$ is a stopping time for a sequence of
\emph{i.i.d.} random variables
$\left\{\widehat{a}[\tau]-\widehat{c}_1[\tau]\right\}_{\tau\in\mathbb{N}}$.
According to Wald's identities (see Theorem 3 in page 488 in
\cite{shiryaev1996probability}), we have, for any $0\leq q<1$,
\begin{align}
&\mathbb{E}\left[Y(q)-q-\widehat{T}(q)\mu\right]=0,\nonumber\\
&\mathbb{E}\left[\left(Y(q)-q-\widehat{T}(q)\mu\right)^2-\widehat{T}(q)\sigma^2\right]=0.\nonumber
\end{align}
It follows directly that, for any $0\leq q<1$
\begin{align}
\mathbb{E}\left[\widehat{T}(q)\right]&=\frac{\mathbb{E}[Y(q)]-q}{\mu},\label{ET_condition}\\
\mathbb{E}\left[\widehat{T}(q)^2\right]&\stackrel{(a)}{=}\frac{\mathbb{E}[Y(q)]-q}{\mu^3}\sigma^2\notag\\
&+\frac{2\mathbb{E}\big[(Y(q)-q)\widehat{T}(q)\big]}{\mu}\notag\\
&-\frac{\mathbb{E}\left[(Y(q)-q)^2\right]}{\mu^2},\label{ET2_condition}
\end{align}
where in step (a), Equation~(\ref{ET_condition}) is also applied.
From the definition of $\widehat{T}(q)$ in Equation~(\ref{T_hat})
and the fact that $\lambda-1\leq\widehat{a}[t]-\widehat{c}_1[t]\leq
\lambda$ for any $t$, we know that, for every $0\leq q<1$, 1) If
$\widehat{T}(q)=1$, then
$|Y(q)-q|=|\widehat{a}[t]-\widehat{c}_1[t]|\le 1$; 2) If
$\widehat{T}(q)\geq 2$, then $0\leq Y(q)<1$. Therefore, in general,
$|Y(q)-q|\le 1$ for every $0\leq q<1$, which, by combining the fact
that $0\leq\widehat{Q}_1<1$, implies that $-1\leq
Y(\widehat{Q}_1)-\widehat{Q}_1\le 1$. Based on this observation,
Equation~(\ref{ET2_condition}) can be further expressed as
\begin{align}\label{ET2_cond_bounds}
\mathbb{E}\left[\widehat{T}(q)^2\right]&\stackrel{(a)}{\le}\frac{\mathbb{E}[Y(q)]-q}{\mu^3}\sigma^2-\frac{2\mathbb{E}\big[\widehat{T}(q)\big]}{\mu}
\nonumber\\&\stackrel{(b)}{=}\frac{\mathbb{E}[Y(q)]-q}{\mu^3}\sigma^2-\frac{2(\mathbb{E}[Y(q)]-q)}{\mu^2},
\end{align}
where in step (a), the lower bound of $Y(q)-q$ is utilized due to
$\mu<0$; and in step (b), Equation~(\ref{ET_condition}) is applied.
By substituting $q$ with $\widehat{Q}_1$ and taking the expectation
of Equations~(\ref{ET_condition}) and (\ref{ET2_cond_bounds}) with
respect to $\widehat{Q}_1$, Equation~(\ref{ETET2}) is derived.

\section{Proof for Lemma~\ref{lemma_GIGI1}}\label{pf_lemma_5}
Consider the number of ``unseen'' packets $\lfloor Q_i[t]\rfloor$
for receiver $i$.The number of data packets that have entered the
encoder buffer up to time-slot $t$ is $\lfloor A[t]\rfloor=\lfloor
\lambda t \rfloor$. When $\lfloor Q_i[t]\rfloor \geq 1$, receiver
$i$ has at least one ``unseen'' packet. In this case, the service
time for receiver $i$ to see one more packet is \emph{i.i.d.}
geometrically distributed with mean $1/\gamma$. When $\lfloor
Q_i[t]\rfloor = 0$, receiver $i$ needs to wait for another data
packet to enter the encoder buffer before serving it.

%The packet injection process satisfies $A[t]=\lambda t$ for $t=1,2,\cdots$.

We construct a D/Ge/1 queue $Q^G(t)$, in which packets arrive one by
one with a fixed interarrival interval $1/\lambda$ and the service
time of the $m^{\text{th}}$ packet is chosen to be equal to that of
seeing the $m^{\text{th}}$ packet in $\lfloor Q_i[t]\rfloor$, which
is \emph{i.i.d.} geometrically distributed. Both queueing system
initiate from the zero state at time $t=0$, i.e., $Q^G(0)=
Q_i[0]=0$. We will show that
\begin{eqnarray}\label{eq_queue_difference}
\left|Q^G(t)- \lfloor Q_i[ t]\rfloor \right|\leq2
\end{eqnarray}
for all integers $t=0,1,\cdots$.
%$Q^G(t)=\lfloor Q_i[t]\rfloor$ for all integer $t=1,2,\cdots$.

First, the $m^{\text{th}}$ packet arrives at
$t=\left\lceil{m}/{\lambda}\right\rceil$ for $\lfloor
Q_i[t]\rfloor$, where $\left\lceil y\right\rceil$ the minimum
integer no smaller than $y$. And the $m^{\text{th}}$ packet arrives
at $t={m}/{\lambda}$ for the constructed D/Ge/1 queue $Q^G(t)$. The
time difference between the two arrival instants satisfies
\begin{eqnarray}\label{eq_arrival_difference}
\left|\left\lceil\frac{m}{\lambda}\right\rceil-\frac{m}{\lambda}\right|<1,~\forall~m=1,2,\cdots.
\end{eqnarray}

%the number of arrived packets of the D/Ge/1 queue $Q^G(t)$ up to any integer time $t$ is $A^G(t)=\lfloor \lambda t \rfloor$, which is equal to that of $\lfloor Q_i[t]\rfloor$.

%since $Q^G(t)$ is a continuous-time queueing system, it may start to serve the next packet at a non-integer time. On the other hand, $\lfloor Q_i[t]\rfloor$ is a discrete-time queueing system, its service for each packet starts from an integer time.

Second, let $T_{s,m}\in \mathbb{N}$ be the service duration of the $m^{\text{th}}$ packet in both queueing systems, $s^G_{m}\in \mathbb{R}$ and $s_m\in \mathbb{N}$ be the service starting instants of the $m^{\text{th}}$ packets for $Q^G(t)$ and $\lfloor Q_i[t]\rfloor$, respectively. We needs to show that
\begin{eqnarray}\label{eq_depart_difference}
|s^G_{m}-s_m|<1,~\forall~m=1,2,\cdots.
\end{eqnarray}
The queueing system of the ``unseen'' packets $\lfloor
Q_i[t]\rfloor$ satisfies
\begin{eqnarray}\label{eq_Q1_update}
s_{m+1} = \max
\left\{\left\lceil\frac{m+1}{\lambda}\right\rceil,s_{m} +
T_{s,m}\right\},
\end{eqnarray}
and the D/Ge/1 queue satisfies
\begin{eqnarray}\label{eq_Q2_update}
s^G_{m+1} = \max \left\{\frac{m+1}{\lambda},s^G_{m} +
T_{s,m}\right\}.
\end{eqnarray}
Using Equations~\eqref{eq_arrival_difference}, \eqref{eq_Q1_update}, and \eqref{eq_Q2_update},
one can prove Equation~\eqref{eq_depart_difference} by induction.

Since the interarrival interval $1/\lambda$ and the packet service duration $T_{s,m}$ of the constructed D/Ge/1 queue $Q^G(t)$ are both no smaller than 1, we have $|Q^G(t+1)-Q^G(t)|\leq1$ for all real $t$. This, together with Equations~\eqref{eq_arrival_difference} and~\eqref{eq_depart_difference}, implies Equation~\eqref{eq_queue_difference}.

According to Proposition 9 in \cite{glynn1993logarithmic}, the constructed
D/Ge/1 queue $Q^G[t]$ satisfies
\begin{eqnarray}
-\lim_{k\to\infty}\frac{1}{k}\mathbb{P}(Q^G>k)=\eta,
\end{eqnarray}
where $\eta$ is defined in Equation~(\ref{eq:rate_queue}). Combining
this with Equation~\eqref{eq_queue_difference}, the asserted
statement follows.

\section{Proof for Lemma~\ref{lemma_OmegaW}}\label{pf_lemma_6}
We have shown step one takes at most $ {K}_jW[t_1^j]$ operations in
decoding interval $ {T}_j$. By Equation~(\ref{Wt_bounds}) and
$Q_1[t_1^j]<1$,
\begin{align}
W[t_1^j]&\le\max_{1\leq i\leq n}Q_i[t_1^j]+1
\le\max_{2\leq i\leq n}Q_i[t_1^j]+2.
\end{align}

By Equations~(\ref{K_bounds}) and \eqref{T_power_def}, ${K}_j$ is
uniquely determined by $Q_1[t_1^j]$ and $ {T}_j$,
$\Big\{\big\{Q_i[t_i^j]\big\}_{1\le i\le n}, {K}_j\Big\}_j$ is a
Markov renewal process.
%Together with the definition of $\widehat{T}$ (Equation~\eqref{T_hat}), the number of packets decoded in the steady state, denoted as $\widehat{K}$ can be derived as
%\begin{align}\label{K_def}
%\widehat{K}=\lfloor \widehat{Q}_1[0]+\widehat{T}\lambda\rfloor.
%\end{align}
Take $\left(\max_{2\leq i\leq n}Q_i[t_1^j]+2\right) {K}_j$ as the
reward gained for $ {K}_j$. Let $\widehat{Q}_i$ denote a random
variable that has the same distribution as the stationary
distribution of $\{Q_i[t_1^j]\}_j$. According to Markov renewal
reward theory (see Theorem 11.4 \cite{ccinlar1975exceptional}), the
average number of operations taken by step one is bounded by
\begin{align}\label{part1_renewal}
\overline{\Omega}_{n,1}&\le\lim_{J\to\infty}\frac{\sum_{j=1}^J\left(\max_{2\leq
i\leq n}Q_i[t_1^j]+2\right) {K}_j}{\sum_{j=1}^J
 {K}_j}\nonumber\\
 &\overset{(a)}{=}\mathbb{E}\left[\mathbb{E}\left[\max_{2\leq i\leq n}\widehat{Q}_i\Big|\left\{\widehat{Q}_i, 2\leq i\leq n\right\}\right]\right]+2\notag\\
 &=\mathbb{E}\left[\max_{2\leq i\leq n}\widehat{Q}_i\right]+2,
\end{align}
where step (a) holds because $K_j$ only depends on $Q_1[t_1^j]$ and is independent of $\{Q_i[t_1^j],2\leq i\leq n\}$.

From the evolution of $Q_i$ shown in Equation~(\ref{rw_normal}), we
know that
\begin{align}
Q_i[t_1^{j}]-t \leq Q_i[t_1^{j}+t], \notag
\end{align}
which yields,
\begin{align}
&T_j\left(\max_{2\leq i\leq n} Q_i[t_1^j]-T_j\right) \leq \sum_{t=1}^{T_j}\max_{2\leq i\leq n}Q_i[t_1^{j-1}+t],\notag
\end{align}
which further implies,
\begin{align}
&\lim_{J\to\infty}\frac{\sum_{j=1}^JT_j\left(\max_{2\leq i\leq n} Q_i[t_1^j]-T_j\right)}{\sum_{j=1}^J T_j}\notag\\
\leq &\lim_{J\to\infty}\frac{\sum_{t=1}^{\infty}{\bf 1}_{\left\{t<\sum_{j=1}^J T_j\right\}}\max_{2\leq i\leq n}Q_i[t]}{\sum_{j=1}^J T_j}\notag\\
=&\mathbb{E}\left[\max_{2\leq i\leq n}Q_i\right], \label{TjMaxQi}
\end{align}
where $Q_i$, defined in Appendix \ref{sec:pftheorem3}, is a random
variable with a distribution as the steady state distribution of
$Q_i[t]$.

Next, we shift our focus to a different Markov renewal process
$\big\{\big\{Q_i[t_i^j]\big\}_{1\le i\le n}, {T}_j\big\}_j$.
According to Markov renewal reward theory (see Theorem 11.4
\cite{ccinlar1975exceptional}), the left hand side of
Equation~(\ref{TjMaxQi}) can be further expressed as
\begin{align}\label{sample_bridge}
&\lim_{J\to\infty}\frac{\sum_{j=1}^JT_j\left(\max_{2\leq i\leq n} Q_i[t_1^j]-T_j\right)}{\sum_{j=1}^J T_j}\notag\\
=&\frac{\mathbb{E}\left[\mathbb{E}\left[\widehat{T}(\widehat{Q}_1)\max_{2\leq i\leq n}\widehat{Q}_i-\widehat{T}(\widehat{Q}_1)^2\Big|\left\{\widehat{Q}_i,1\leq i\leq n\right\}\right]\right]}{\mathbb{E}\left[\mathbb{E}\left[\widehat{T}(\widehat{Q}_1)\Big|\left\{\widehat{Q}_i,1\leq i\leq n\right\}\right]\right]}\notag\\
=&\mathbb{E}\left[\max_{2\leq i\leq
n}\widehat{Q}_i\right]-\frac{\mathbb{E}[\widehat{T}^2]}{\mathbb{E}[\widehat{T}]},
\end{align}
Comparing Equation~\eqref{TjMaxQi} and
Equation~\eqref{sample_bridge}, we have
\begin{align}\label{sample_upper}
\mathbb{E}\left[\max_{2\leq i\leq
n}\widehat{Q}_i\right]\le\mathbb{E}\left[\max_{2\leq i\leq
n}Q_i\right]+\frac{\mathbb{E}[\widehat{T}^2]}{\mathbb{E}[\widehat{T}]},
\end{align}
which, by combining with Equation~(\ref{part1_renewal}), yields
\begin{align}
\overline{\Omega}_{n,1}&\leq \mathbb{E}\left[\max_{2\leq i\leq
n}Q_i\right]+\frac{\mathbb{E}[\widehat{T}^2]}{\mathbb{E}[\widehat{T}]}+2\notag\\
&\stackrel{(a)}{\le}\overline{W}_n+\frac{\mathbb{E}[\widehat{T}^2]}{\mathbb{E}[\widehat{T}]}+3,
\end{align}
where in step (a), Equation~\eqref{EW_bounds} is applied.

According to Lemma~\ref{lemma_Y}, the second term in the above
equation is independent of the number of receivers $n$, and thus the
proof is complete.

\section{Proof for Lemma~\ref{lemma_Omega}}\label{pf_lemma_7}
Note that,
\begin{align}
\overline{\Omega}_{n,2}&\le\lim_{J\to\infty}\frac{\sum_{j=1}^J
M_\text{C}( {K}_j)^3}{\sum_{j=1}^J
 {K}_j}\nonumber\\&\stackrel{(a)}{\le}\lim_{J\to\infty}\frac{\sum_{j=1}^J
M_\text{C}(\lambda {T}_j+1)^3}{\sum_{j=1}^J
 {T}_j}\cdot\lim_{J\to\infty}\frac{\sum_{j=1}^{J}
{T}_j}{\sum_{j=1}^{J}
{K}_j}\nonumber\\
&\stackrel{(b)}{=}\frac{\mathbb{E}\left[\mathbb{E}\left[M_\text{C}(\lambda
\widehat{T}(\widehat{Q}_1)+1)^3\Big|\widehat{Q}_1\right]\right]}{\mathbb{E}\left[\mathbb{E}\left[\widehat{T}(\widehat{Q}_1)|\widehat{Q}_1\right]\right]}\cdot\frac{1}{\lambda}\nonumber\\
&=\frac{M_\text{C}\mathbb{E}\left[\mathbb{E}\left[(\lambda
\widehat{T}(\widehat{Q}_1)+1)^3\Big|\widehat{Q}_1\right]\right]}{\lambda
\mathbb{E}\left[\widehat{T}\right]},
\end{align}
where in step (a), Equation~(\ref{K_bounds}) is used to bound $
{K}_j$, and in step (b), Markov renewal reward theory (see Theorem
11.4 \cite{ccinlar1975exceptional}) is applied on the first limit
and the second limit has been given by Equation~(\ref{lim_trivial}).

\section{Proof for Lemma~\ref{lemma_Omega_lower}}\label{pf_lemma_8}
Note that the number of operations to decode the $K_j$ packets in
the $j^\text{th}$ decoding interval $T_j$ is lowered bounded by the
number of nonzero elements in the $K_j\times\left(W[t_1^j]+
K_j\right)$ decoding matrix. From
Equations~\eqref{At_def}\eqref{eq_W}\eqref{Zt_2} and \eqref{Zt_1},
$|W[t+1]-W[t]|\le 1, \forall t$. Thus, there are at least
$W[t_1^j]-T_j$ nonzero elements in each rows of the decoding matrix.
The complexity taken to decode the $K_j$ packets is at least
$K_j(W[t_1^j]-T_j)$. By Equation~(\ref{Wt_bounds}) and
$Q_1[t_1^j]\ge 0$,
\begin{align}\nonumber
W[t_1^j]&\ge\max_{1\leq i\leq n}Q_i[t_1^j]-1 \ge\max_{2\leq i\leq
n}Q_i[t_1^j]-1.
\end{align}
With the above facts, we can derive the lower bound of the average
decoding bound as
\begin{align}\label{lower_renewal}
\overline{\Omega}_n&\ge\lim_{J\to\infty}\frac{\sum_{j=1}^J\left(\max_{2\leq
i\leq n}Q_i[t_1^j]-1-T_j\right) {K}_j}{\sum_{j=1}^J
 {K}_j}\nonumber\\
 &\overset{(a)}{=}\mathbb{E}\left[\max_{2\leq i\leq n}\widehat{Q}_i\right]-\lim_{J\to\infty}\frac{\sum_{j=1}^J\left(T_j\right) {K}_j}{\sum_{j=1}^J
 {K}_j}-1\notag\\
 &\overset{(b)}{\ge}\mathbb{E}\left[\max_{2\leq i\leq n}\widehat{Q}_i\right]-\lim_{J\to\infty}\frac{\sum_{j=1}^J\left(T_j\right)(\lambda{T}_j+1)}{\sum_{j=1}^J
 {T}_j}\frac{1}{\lambda}-1\nonumber\\
 &\overset{(c)}{=}\mathbb{E}\left[\max_{2\leq i\leq
 n}\widehat{Q}_i\right]-\lambda\frac{\mathbb{E}[\widehat{T}^2]}{\mathbb{E}[\widehat{T}]}-\frac{\lambda+1}{\lambda},
\end{align}
where step (a) uses the same argument in
Equation~\eqref{part1_renewal}, in step (b)
Equations~\eqref{lim_trivial} and \eqref{K_bounds} are directly
applied, and step (c) uses the same argument in
Equation~\eqref{sample_bridge}.

From the evolution of $Q_i$ shown in Equation~(\ref{rw_normal}), we
know that
\begin{align}
Q_i[t_1^{j}]+t \ge Q_i[t_1^{j}+t], \notag
\end{align}
which yields,
\begin{align}
&T_j\left(\max_{2\leq i\leq n} Q_i[t_1^j]+T_j\right) \ge
\sum_{t=1}^{T_j}\max_{2\leq i\leq n}Q_i[t_1^{j-1}+t],\notag
\end{align}
which, by following the similar deductions in
Equations~\eqref{TjMaxQi} and \eqref{sample_bridge} leads to
\begin{align}\label{sample_lower}
\mathbb{E}\left[\max_{2\leq i\leq
n}\widehat{Q}_i\right]\ge\mathbb{E}\left[\max_{2\leq i\leq
n}Q_i\right]-\frac{\mathbb{E}[\widehat{T}^2]}{\mathbb{E}[\widehat{T}]},
\end{align}
which, combining with Equation~\eqref{lower_renewal}, yields
\begin{align}
\overline{\Omega}_n&\ge\mathbb{E}\left[\max_{2\leq i\leq
n}Q_i\right]-(1+\lambda)\frac{\mathbb{E}[\widehat{T}^2]}{\mathbb{E}[\widehat{T}]}-\frac{\lambda+1}{\lambda}\nonumber\\
&\stackrel{(a)}{\ge}\overline{W}_{n-1}-(1+\lambda)\frac{\mathbb{E}[\widehat{T}^2]}{\mathbb{E}[\widehat{T}]}-\frac{2\lambda+1}{\lambda},
\end{align}
where in step (a), Equation~\eqref{EW_bounds} is applied.

According to Lemma~\ref{lemma_Y}, the second and third terms in the
above equation are independent of the number of receivers $n$, and
thus the proof is complete.
 \fi
\else \fi

{\footnotesize
\bibliographystyle{ieeetr}
\bibliography{reference}
}

\end{document}